\documentclass[aps,prb,twocolumn,amsmath,amssymb,nofootinbib,floatfix]{revtex4-1}

\usepackage{amsmath}
\usepackage{amssymb}
\usepackage{amsthm}
\usepackage[dvipdf]{color}
\usepackage{graphicx}
\usepackage{dcolumn}
\usepackage{bm,subfigure} 
\usepackage{xcolor,colortbl}

\DeclareMathOperator{\arcsinh}{arcsinh}
\DeclareMathOperator{\tr}{tr}

\begin{document}
\title{Maxwell plates and phonon fractionalization}

  \author{Kai Sun}
  \author{Xiaoming Mao}

 \affiliation{
 Department of Physics,
  University of Michigan, Ann Arbor, 
 MI 48109-1040, USA
 }

\begin{abstract}
In the past a few years, topologically protected mechanical phenomena have been extensively studied in discrete lattices and networks, leading to a rich set of discoveries such as topological boundary/interface floppy modes and states of self stress, as well as one-way edge acoustic waves.  In contrast, topological states in continuum elasticity  without repeating unit cells  remain largely unexplored, but offer wonderful opportunities for both new theories and broad applications in technologies, due to their great convenience of fabrication.  In this paper we examine continuous elastic media on the verge of mechanical instability, extend Maxwell-Calladine index theorem to continua in the nonlinear regime, classify  elastic media based on whether stress can be fully released, 
and identify two types of elastic media with topological states.  The first type, which we name ``Maxwell plates'', are in strong analogy with Maxwell lattices, and exhibit a sub-extensive number of holographic floppy modes.  The second type, which arise in thin plates with a small bending stiffness and a negative Gaussian curvature, exhibit fractional excitations and topological degeneracy, in strong analogy to $Z_2$  spin liquids and dimerized spin chains.
\end{abstract}

\maketitle

\section{Introduction}
\begin{figure}[t]
	\centering
	\includegraphics[width=.8\columnwidth]{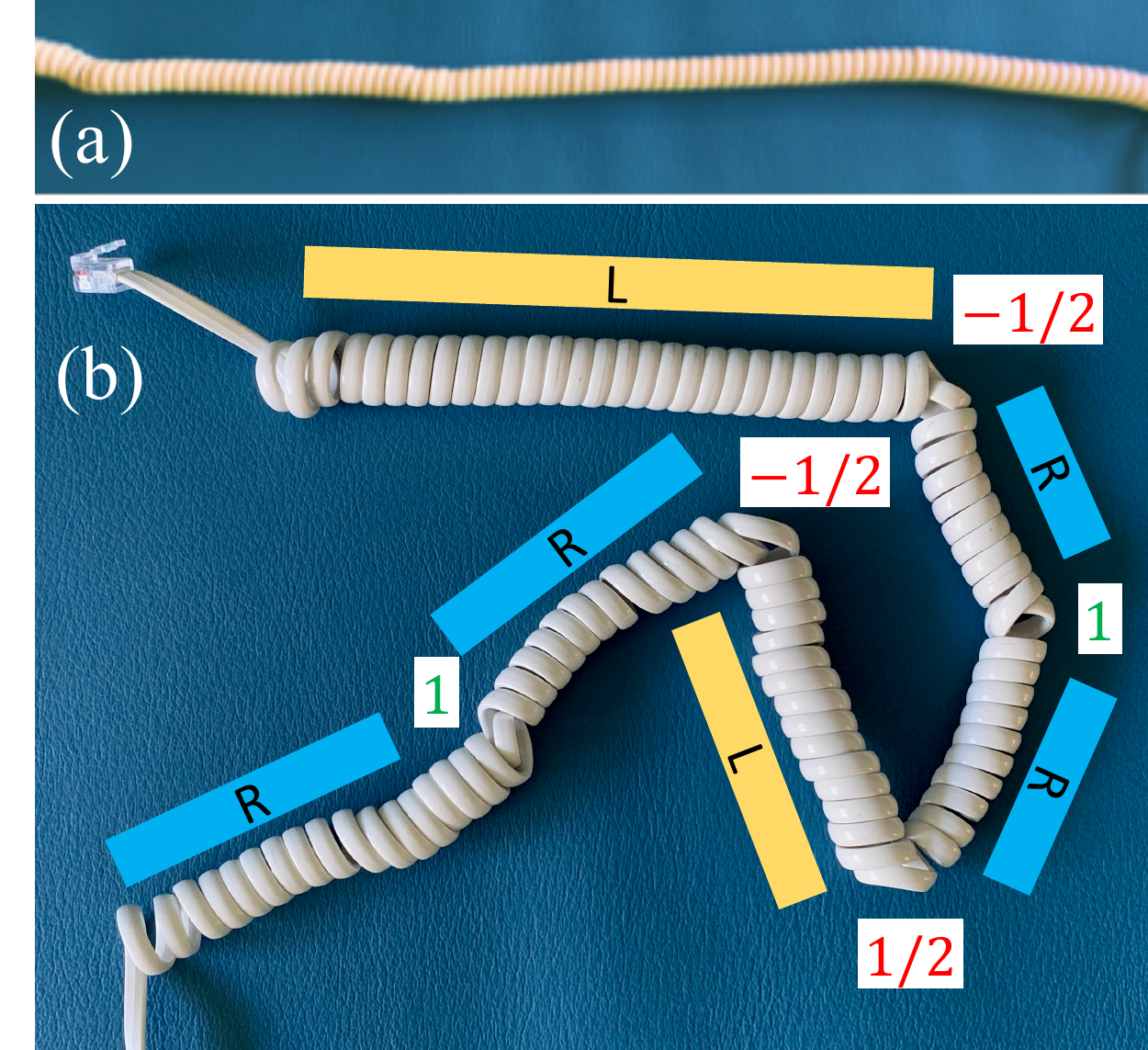}
	\caption{Kinks in helical cords. (a) A kink-free helical cord is straight.  (b) Kinks sharply turn the helical cord.  Chirality (L or R) of domains and topological charge of kinks are marked (see discussions in Secs.~\ref{SEC:MiniSurf} and \ref{SEC:FracRibb}).
	}
	\label{fig:cords}
\end{figure}

On a landline telephone, the headset is usually connected with the phone via a helical cord. Such a helical cord, although initially nice and straight [Fig.~\ref{fig:cords}(a)], often tangles up after use [Fig.~\ref{fig:cords}(b)]. A careful examination would reveal that a tangled helical cord is sharply different from tangled hair or ropes. The later come from knotted structures, while a tangled helical cord is usually knot free. Instead, it is due to kinks, at which a helical cord changes its chirality and turns sharply. 
Where do these kinks come from? Why do they turn a nice and straight cord into a tangled mess?  Why are they so hard to get rid of? 
Interestingly,  answers to these questions, as we show below, have roots in the fascinating physics and mathematics of materials and structures at the verge of mechanical instability, to which these helical cords belong.

Mechanical stability is concerned with the response of a material to external loads---whether it holds its shape like a solid or flows like a liquid.  At the verge between these two types of behaviors, interesting physics arises, such as critical phenomena~\cite{Jacobs1995,Liu2010,Ellenbroek2011,Ellenbroek2014,Zhang2017} and topological states~\cite{Kane2014,Lubensky2015,Paulose2015a,Stenull2016,mao2018maxwell,Zhou2018}.  One peculiar feature of these materials is that they can exhibit a small number of floppy modes, i.e., normal modes of deformation that cost little elastic energy, but at the same time remain stable under other types of loads.  By controlling these floppy modes, which are the low energy excitations of the system, mechanical response of a material can be precisely programmed, which has broad applications such as switchable, actuatable, deployable materials~\cite{bertoldi2017flexible,Mullin2007,schenk2013geometry,Florijn2014,Paulose2015,filipov2015origami,Chen2015,dudte2016programming,Rocklin2017,tang2017programmable}.  

Among these fascinating systems right at the verge of mechanical instability, a subcategory, known as Maxwell networks/lattices~\cite{Lubensky2015}, has been extensively studied recently, and provided deep insight into many problems in soft matter~\cite{Mao2010,Mao2011a,Broedersz2011,Mao2013b,Mao2013c,Dennison2013,Mao2015,Zhang2015a,Zhang2016,Zhang2017,Liarte2019}, metamaterials~\cite{Chen2014,Paulose2015,Rocklin2017,Zhang2018,Ma2018} and mechanobiology~\cite{Alvarado2013,Feng2015,Feng2016,Sharma2016,Ronceray201514208,feng2019cell}. 
Thanks to the discreteness of the degrees of freedom and constraints in these lattices/networks, 
within linear elasticity, a universal theorem, known as the the Maxwell-Calladine index theorem~\cite{Calladine1978,PellegrinoCal1986,Kapko2009,Sun2012} can be rigorously proved, which reveals a profound relation between floppy modes and states of self stress.
States of self stress, as we define rigorously below,  describes modes of distributing stress on  a network leaving all components in force balance and thus characterize the ability of a network to carry stress.  In contrast to floppy modes which describe kinematics of a structure, states of self stress focus on the statics.
In addition to demonstrating a universal connection between these two seemingly unrelated quantities, this theorem also played a crucial role in the discovery of topological mechanics in Maxwell lattices and networks~\cite{Kane2014,Lubensky2015}. Moreover, there is a rigorous duality relation between the geometry of  floppy modes and states of self stress via Maxwell reciprocal diagrams~\cite{maxwell1864xlv,crapo1994spaces,mitchell2016mechanisms,Zhou2019}.

In contrast, the verge of mechanical instability in continuous elastic media is much less understood, especially in terms nonlinear elasticity which naturally arises when low dimensional elastic objects are embedded in a higher dimensional space~\cite{Witten2007,Klein1116,sharon2010mechanics,Efrati2011,EFRATI2009762,Armon1726,Kim1201,chen2012nonlinear,guo2014shape,gladman2016biomimetic}.  It is worth pointing out that here mechanical instability refers to intrinsic instability coming from material properties and geometry, rather than instabilities caused by external stress, such as buckling.  In a nonlinear continuous elastic medium, is there a connection between floppy modes and states of self stress, similar to that in lattices/networks? If so, does it lead to states with nontrivial topological features?

In continuum, floppy modes correspond to vanishing elastic moduli for certain types of load,  
whereas states of self stress dictate the possibility for an elastic body to carry residual stress. 
Stress can be generated in elastic bodies in many different ways, such as frozen in stress from manufacturing, topological defects in crystals, and inhomogeneous growth in biological tissue~\cite{Alexander1998,Withers2001,goriely2007definition}.  When the elastic body is allowed to relax in absence of external load, these stresses may or may not be fully released, and the stress that remains becomes residual stress.  Physically, the existence of residual stress implies that if we cut the system into two parts, neither of them will retain its original shape. Instead, the two pieces will both deform to reduce elastic energy associated with the residual stress. Residual stress often results in ``weak points'' for mechanical failure, but they are also very useful in toughening brittle materials and can be introduced intentionally.  
Stress-free solid materials, on the contrary, usually requires fine tuning, e.g., through careful annealing. Once achieved, cutting the material into pieces will not affect its shape and/or spacial configuration, in contrast to stressed materials.

In this manuscript, we examine floppy modes and states of self stress in continuum elasticity, and demonstrate that Maxwell's counting argument still applies, and surprisingly, even in the nonlinear regime.  Based on Maxwell's counting, we classify elastic bodies at different dimensions and study their ground states and low-energy excitations. In light of the counting argument and insights gained from the study of quantum topological states, we identify two types of  thin elastic objects, where topological floppy modes/low-energy excitations arise. The first class is called Maxwell plates. Such a plate is stress-free and contains sub-extensive number of holographic floppy modes, in strong analogy to 2D Maxwell/isostatic lattices. The second class is also stress-free and features only one floppy mode. In addition, low-energy excitations in these system are fractional particles, in strong analogy to a $Z_2$  spin liquid and/or a dimerized spin chain. The nature of these fractional excitations and their connection with topological degeneracy are also discussed. In addition to common features of fractional excitations, fractional excitations in these elastic systems show holographic feature, i.e., the location and configuration of a bulk fractional excitation are be fully dictated by the edge. Interestingly, these topological and holographic features naturally answer the three questions about kinks in a telephone cord mentioned above.

\section{Maxwell lattices and networks}\label{SEC:counting}
In this section, we provide a brief review about elasticity in discrete systems, e.g., Maxwell lattices and networks. Mechanical properties of discrete networks can be conveniently analyzed via a counting argument due to Maxwell~\cite{Maxwell1864,Calladine1978,Sun2012}.  

The idea of Maxwell's counting is based on a simple insight about the interplay between degrees of freedom and constraints. If the number of constraints $N_c$ exceeds the
number of the degrees of freedom $N_{\textrm{d.o.f.}}$, the system is often ``rigid''. On the contrary,
if $N_c<N_{\textrm{d.o.f.}}$, the system is often ``floppy''. In this picture, the verge of rigidity is marked by the Maxwell condition
\begin{align}
N_{\textrm{d.o.f.}}=N_{c},
\label{eq:z2D}
\end{align}
and lattices satisfying this condition are called ``Maxwell lattices''. This Maxwell condition also plays an important role in a 
wide range of elastic systems at the verge of mechanical stability~\cite{Liu2010,Mao2010,Broedersz2011,During2013,Kane2014}.

Before applying to elasticity,  here we first demonstrate these ideas by considering a set of linear equations, where $N_{\textrm{d.o.f.}}$ is just the number of variables and 
$N_{c}$ is the number of equations, because each equation enforces one linear constraint on these variables.
It is easy to realize that the number of  ``un-fixed'' degrees of freedom can be obtained as
\begin{align}
N_{\textrm{un-fixed d.o.f.}} = N_{\textrm{d.o.f.}}-N_{c}+N_{\textrm{r.c.}}
\label{eq:N0}
\end{align}
Here, because constraints may not all be independent of one another, redundant constraints needs to be removed in this counting 
and  $N_{\textrm{r.c.}}$ here represent the number of redundant constraints.
For linear equations, this counting relation can be rigorously proved. However, it must be emphasized that when nonlinearity are taken into account, such
a counting argument is not generically expected, due to the complicated nature of nonlinear constraints. 

For discrete lattices/networks, the counting relation Eq.~\eqref{eq:N0}  can be applied if we focus on linear response around stress-free ground states.
Here, it is easy to realize that the number of  ``un-fixed'' degrees of freedom $N_{\textrm{un-fixed d.o.f.}}$ gives the number of zero modes $N_0$, i.e. normal modes that don't
cost any elastic energy. As for redundant constraints, they have a one-to-one correspondence with states of self stress and thus $N_{\textrm{r.c.}}$ must 
coincide with the number of states of self stress $N_s$.
This is because  by definition a redundant constraint tries to fix a degree of freedom which is already pinned by other constraints. Thus, in order to keep the network stress-free, this redundant constraint must be ``fine-tuned'' such that it is compatible with other constraints. If not compatible, such a redundant constraint will result in conflict constraints and thus generates residual stress,   giving rise to a state of self-stress. 
As a result, the counting argument discussed above leads to the Maxwell-Calladine index theorem.
\begin{align}
N_0 -N_{s} = N_{\textrm{d.o.f.}}-N_{c}
\label{eq:N0s}
\end{align}
which shows that the number of zero modes $N_0$ and the number of states of self-stress $N_{\textrm{s}}$ are directly related. 
In particular, for networks/lattices at the verge of mechanical instability, i.e., Maxwell lattices/networks with $N_{\textrm{d.o.f.}}=N_{c}$, these two numbers must coincide ($N_0 =N_{s}$).

\begin{figure}[t]
	\centering
	\includegraphics[width=1\columnwidth]{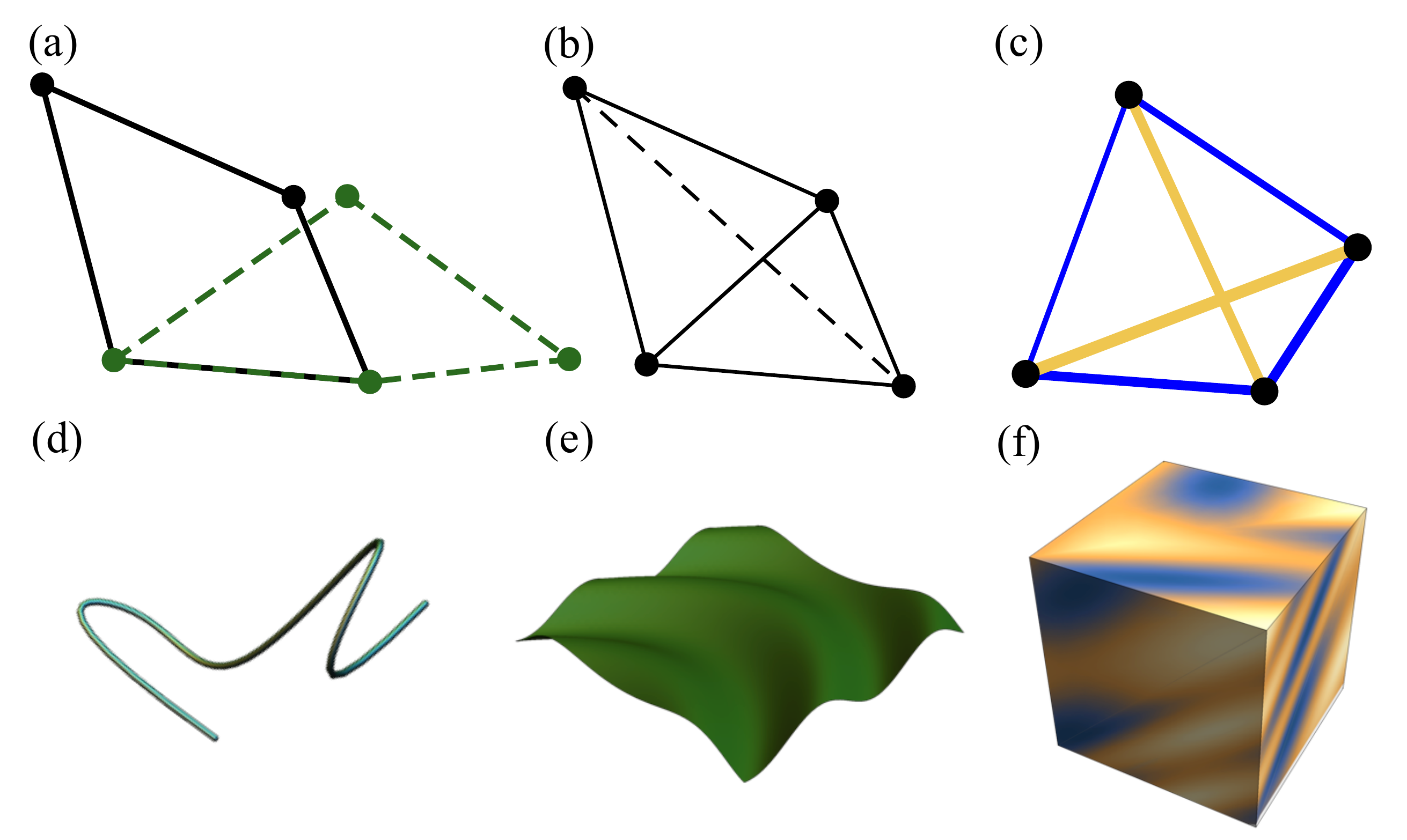}
	\caption{Counting degrees of freedom and constraints.  (a-c) 2D discrete networks composed of sites (black dots) and bonds (black solid lines), which are 
	(a) underconstraint, (b) isostatic and (c) over-constraint. In (a), $N_{\textrm{d.o.f.}}=8, N_{c}=4, N_{s}=0$, so $N_0=4$ and thus the network has one floppy mode (green dashed configuration), i.e., $N_f=1$ according to Eq.~\eqref{eq:Nm}. (b) Introducing one more constraint makes $N_{c}=5$ and $N_f=0$, and thus the network is isostatic and at the verge of mechanical instability.  (The dashed line is not a constraint.)  (c) The addition of more constraints into the network leads to states of self stress: when the new diagonal bond is not at the prescribed length [black dashed line in (b)] the network has residual stress on bonds (yellow for tension and blue for compression). (d-f) $d$ dimensional elastic bodies embedded in a $D=3$ dimensional space.  (d) A curve of $d=1$, with bending and torsional stiffness ignored, is under constrained, in analogy to (a).  (e) A surface of $d=2$, with bending stiffness ignored, is at the verge of mechanical instability, in analogy to (b).  (f) A body of $d=3$ is always over-constrained, in analogy to (c), and able to host residual stress (blue and yellow colors).}
	\label{fig:counting}
\end{figure}

In Fig.~\ref{fig:counting}(a-c), we demonstrate the Maxwell-Calladine theorem by considering a network consisting of point-like sites (hinges) connected by central-force bonds (struts) and assume that all bonds are at their rest-lengths in the reference state. Here, each site has $D$ degrees of freedom, where $D$ is the spatial dimension, and each bond provides one constraint. Thus the Maxwell-Calladine theorem takes the form of 
\begin{align}
N_0-N_{s}= N D - N_b   = ND - N\langle z \rangle/2  .
\label{eq:QC}
\end{align}
Here $N D$ is the total number of degrees of freedom $N_{\textrm{d.o.f.}}$, with $N$ being the number of sites.
The total number of constraint $N_{\textrm{c}}$ is now given by the number of bonds $N_b$, which can be written as $N \langle z \rangle/2$ where $\langle z \rangle$ is the average coordination number (number of bonds connecting to one site) of the network. 
For a periodic lattice, the Maxwell condition [Eq.~\eqref{eq:z2D}] here takes the form of $\langle z \rangle =2D$.
In Fig.~\ref{fig:counting}(c), we present one example demonstrating the relation between redundant constraints and states of self stress. When the last bond is added to the network, if it is not fine-tuned to the length prescribed by other bonds, residual stress is introduced.  Analogous continuous elastic media (d-f) are discussed in Sec.~\ref{SEC:ContinuumCounting}.

One important feature of Maxwell lattices/networks lies in its mechanical properties under open boundary conditions. An infinitely large Maxwell lattice is perfectly constrained everywhere (unless some constraints are redundant).  When a finite piece of Maxwell lattice is cut off, under open boundary conditions, the lattice will have a subextensive number of zero modes, coming from the cut bonds on the surface, inducing a deficit of constraints.  These zero modes exhibit very interesting properties, and their localization has been shown to be a topologically protected property, dictated by the bulk phonon structure~\cite{Kane2014,Lubensky2015}.

We conclude this section by discussing the relation between the Maxwell condition [Eq.~\eqref{eq:z2D}] and isostaticity. For an elastic system, if is often useful 
to exclude trivial degrees of freedom associated with rigid translations and rotations of the whole system, and thus
the number of floppy modes is defined as
\begin{align}
N_f = N_0 - \frac{D(D+1)}{2} ,
\label{eq:Nm}
\end{align}
which describes elastic deformations that cost no elastic energy. Networks that satisfy $N_f=N_s=0$ are called ``isostatic'' structures, which are both stable (no floppy modes) and stress-free [See Fig.~\ref{fig:counting}(b) for an example]. According to the Maxwell-Calladine theorem, isostatic systems should satisfy the relation $N_{\textrm{d.o.f.}}-D(D+1)/2=N_{c}$. This relation differs slightly from the 
Maxwell condition [Eq.~\eqref{eq:z2D}], due to the extra term $D(D+1)/2$ which comes from excluding trivial degrees of freedom (i.e. rigid body translations/rotations).
This difference often becomes negligible for large systems or in the continuum, where $N_{\textrm{d.o.f.}}$ and $N_c$ diverge.  

\section{Continuous elasticity of non-flat media}
In this section, we write down the elastic theory for a $d$-dimensional elastic medium embedded in a $D$-dimensional Euclidean space ($d \le D$). In this convention, the typical 3D elastic theory corresponds to $d=D=3$, while a 2D plate embedded in the 3D space has $d=2$ and $D=3$.
For 2D plates, unless stated otherwise, in this manuscript, we focus on plates that are homogenous along the thickness direction 
(e.g. a plate formed by identical 2D layers stacked vertically on top of each other), and thus the two sides of each plate are always fully equivalent, in contrast to elastic shells. 
It turns out that this symmetry plays an important role in the fractional excitations we discuss below.

The spatial configuration of a $d$-dimensional object in a $D$-dimensional Euclidean space is described by the mapping $\mathbf{r} \to \mathbf{R}$, where $\mathbf{r}$ and $\mathbf{R}$ are the $d$- and $D$-dimensional coordinates of the original space and the target space respectively. For $d=D$, the elastic energy depends on the  deformation of the elastic medium, characterized by the metric tensor (defined below). For a homogenous or nearly homogenous system (i.e., elastic moduli remain constants but stress can slowly vary in space), the elastic energy takes the following functional form
\begin{align}
E_s=\int dr \sqrt{\det g_0} \left\{ \frac{B-\frac{2}{d}G}{2} \tr(g-g_0)^2+ G \tr[(g-g_0)^2]\right\}
\label{eq:Es}
\end{align}
where $B$ ($G$) is the bulk (shear) modules  and $g$ is the metric tensor
$g_{\alpha,\beta}=\partial_\alpha \mathbf{R}\cdot \partial_\beta \mathbf{R}$,
also known as the first fundamental form, with $\partial_\alpha=\partial/\partial \mathbf{r}_\alpha$ ($\alpha=1$, $2$, $\ldots$, $d$). Same as $g$, $g_0$ is also a symmetric rank-2 tensor ($d\times d$), which is usually called the reference metric tensor. As shown in Eq.~\eqref{eq:Es}, $g_0$ is the target for $g$, if we try to minimize the elastic energy.  In terms of discrete elastic networks defined in Sec.~\ref{SEC:counting}, $g=g_0$ corresponds to the configuration where all bonds are at their rest length.

When $g_0=I$, the conventional $D$-dimensional elastic theory is recovered with the strain tensor $\epsilon=g-I$. In this manuscript, we consider more generic cases where $g_0$ may deviate from $I$. For $d=D$, such a generic $g_0$ describes prestressed materials, in which $g_0$ can never be reduced to identity in any coordinate systems. 
From Eq.~\eqref{eq:Es}, it is easy to realize that for $g_0 \ne I$, deforming a system (i.e. making $\mathbf{R}\ne \mathbf{r}$ and thus $g\ne I$) may actually reduce the elastic energy, which corresponds to fully or partially releasing of build-in stress via deformation.

For $d<D$ (e.g. $d=2$ and $D=3$), there is one more reason to consider a generic $g_0$ beyond identity: if the ground state of the 2D plate is curved, $g_0$ may not be reduced to identity in any coordinate systems, and thus it is necessary to explore this more generic $g_0 \ne I$ case. For $d<D$, e.g., a 2D sheet embedded in 3D, the elastic energy can be separated into two parts
\begin{align}
E=E_s+E_b.
\label{eq:d=2D=3}
\end{align}
where the first term $E_s$ describes in plane stress while the second term $E_b$ describes the energy cost associated with bending.
(For even lower $d<2$ or higher $D>3$, more terms can arise, such as the torsion term, which can be treated along the same line but will not be considered explicitly here).
For $d=2$ and $D=3$, this elastic energy [Eq.~\eqref{eq:d=2D=3}] can be obtained by considering a thin 3D plate with thickness $h$, whose elastic energy takes the form of Eq.~\eqref{eq:Es}~\cite{willmore1993riemannian,Klein1116}. 
When the thickness of the plate $h$ is small, the total elastic energy can be written as a power-law expansion of $h$. By keeping the leading order terms, Eq.~\eqref{eq:d=2D=3} is obtained with $E_s \propto h$ and $E_b \propto h^3$, while higher order terms (e.g. $O(h^5)$) will be ignored in this manuscript. For the special case of a flat plate, this construction is demonstrated in Landau's book~\cite{Landau1986}. Here, we consider more generic 2D plates which are not flat.

The first term of Eq.~\eqref{eq:d=2D=3} takes the same form as Eq.~\eqref{eq:Es} but with two modifications
\begin{align}
E_s=h \int dr{\sqrt{\det g_0}}\large\{ \frac{B_0-G_0}{2} & \tr(g-g_0)^2
\nonumber\\
&+ G_0 \tr[(g-g_0)^2]\large\}
\label{eq:Es_2D}
\end{align}
First, this $E_s$ has an extra pre-factor $h$, which is the thickness of the plate, and secondly, $g$ and $g_0$ are now $d\times d$ matrices, instead of $D\times D$.
The elastic moduli here are $B_0=\frac{9 B G}{4(3B+4G)}$ and $G_0=G/4$. 

For plates that are homogenous along the thickness direction, deformations that make the plate non-flat will cost bending energy, which is the $E_b$ term here. Mathematically, bending is described by the so-called the second fundamental form, and thus $E_b$ shall be in general a functional of the second fundamental form. 
As will be shown below, for 2D plates, it is always possible to reach $g=g_0$ at least locally, and thus, for a homogenous or nearly-homogenous 2D plate, 
$E_b$ only depends on two qualities, the mean and Gaussian curvatures ($H$ and $K$)
\begin{align}
E_b=h^3  \int dr{\sqrt{\det g_0}}\frac{G}{12}\left[ \frac{8(3B+G)}{3B+4G} H^2 -2 K \right]
\label{eq:Eb}
\end{align}
which scales as $h^3$. For a flat plate, this bending energy recovers the text-book example of the elastic theory of thin plates~\cite{Landau1986}.

It is worthwhile to emphasize that because although the $E_b$ term prefers a plate to be flat, the $E_s$ term may prefer a non-flat ground state, if $g_0$ gives a non-zero curvature, as we discuss below.

For thin plates (small $h$), because $E_s\propto h$ while $E_b\propto h^3$, $E_b \ll E_s$. Thus, $E_s$ dominates the elastic energy and $E_b$ can be treated as a small perturbation. 
In the rest of the manuscript, we will follow this perturbative approach and all the conclusion would be accurate to the first order of $E_b$ (i.e., $~h^3$). Higher order contributions, e.g. $~h^5$ will be ignored.

\section{Generalized Maxwell's counting and Janet-Cartan theorem}\label{SEC:ContinuumCounting}

\begin{table*}\label{tab:class}
\centering
\begin{tabular}[c]{| c | c | >{\columncolor[rgb]{0.88,1,1}}c || c | c | c | >{\columncolor[rgb]{0.88,1,1}}c |}
\hline
  &\multicolumn{2}{c||}{Maxwell systems}   	&\multicolumn{4}{c| }{ Over-constrained systems} \\
  \hline
  &Maxwell Lattices 	& Maxwell Plates	&Typically solids	& \multicolumn{3}{ c|}{ Plates with bending stiffness}\\
  & 
  $D=2$ or $3$	& $d=2$, $D=3$	& $d=D=2$ or $3$	& \multicolumn{3}{ c| }{ $d=2$, $D=3$}\\
\hline
Stress free & Yes$^1$  & Yes$^2$   & No unless $g_0$ is flat  & No in general & \multicolumn{2}{ c| }{ Yes if criterion in Sec.~\ref{SEC:PlatBend} is met}\\
\hline
Gaussian curvature & - & Any & - & Any & Non-negative  & Negative \\
\hline
Floppy modes & Subextensive  & Subextensive & None  & None  & None & 1\\
\hline
Holographic & 	Yes	 & Yes 	& No    & No  & No & Yes \\
 \hline
Shapes &  Any & Any & Any & Any & Sphere or flat & Minimal surface \\
 \hline
Fractional excitations &  No & No  & No	& No & No & Yes\\
 \hline
\end{tabular}
\caption{Summary of results. Two types of continuum media with nontrivial topological features are highlighted in cyan. $^1$ if the lattice doesn't contain redundant constraints.  $^2$ if the plate is under open boundary conditions and does not hit singularity points.}
\end{table*}

In this section and the following section, we generalize Maxwell's counting argument to continuum elasticity. Similar to discrete systems, this generalized counting argument provides a simple principle for identifying systems with floppy modes and/or states of self-stress. In addition, we also show that predictions from this generalized Maxwell's counting is in full alignment with the mathematical theorems on local embedding
~\cite{janet1926possibilite,cartan1927possibilite}, global embedding and rigidity.

In this section, we focus on systems with elastic energy $E=E_s$, while more general elastic energy (e.g. with bending $E_b$) will be studied in Sec.~\ref{SEC:PlatBend}.  
Systems with $E=E_s$ include two possible cases. The first one is $d=D$, where $E_s$ is naturally the only term in the elastic energy. The second case is for $d<D$, e.g. a 2D plate in a 3-dimensional space. Here, in general, the elastic energy contains more terms beyond $E_s$, e.g., $E_b$. However, as shown above, when the plate is thin enough, $E_b$ becomes negligible, and thus only $E_s$ needs to considered. 

\subsection{Counting argument and its predictions}
Due to nonlinearity, the counting argument [Eq.~\eqref{eq:N0s}] is not generically expected here. In this section, we first perform the counting analysis while ignoring 
possible problems caused by nonlinearity. These predictions will then be checked in the next section. Surprisingly, as shown below, for all generic cases considered here, 
we found that this counting argument happens to provide correct predictions. Of course, it must be emphasized that because such a coincidence cannot be universally proved,
there could exist special cases where such a relation is invalid.

In a $D$-dimensional space, each real-space point has $D$ degrees of freedom. As for constraints, following the same spirit of Maxwell's counting argument, $E_s$ can be treated
as constraints $g=g_0$ , i.e. the elastic energy vanishes if $g=g_0$, in analogy to the a spring network where all springs are at its rest length. The total number of constraints enforced by this relation is $d(d+1)/2$ per real-space point, because $g=g_0$ is a $d$-dimensional symmetric rank-2 tensor with $d(d+1)/2$ independent variables. In mathematical literature, this quantity $d(d+1)/2$ is often referred to as the Janet dimension.

If $D>d(d+1)/2$ (e.g. a 1D chain in 3D space with $E=E_s$), the system is under-constrained, and thus it must be floppy with extensive number of floppy modes. In general, unless some incompatible constraints happen to arise in the system, we expect no state of self-stress, which implies that in general, the ground state is stress-free and can reach $g=g_0$.

If $D<d(d+1)/2$ (e.g. $d=D>1$), the system is over-constrained and thus we expect extensive number of states of self-stress. These systems don't have to contain floppy modes, and thus in general we expect the system to be rigid with one unique ground state.

As for the marginal case where the spatial dimension and the Janet dimension coincides, $D=d(d+1)/2$, the system is at the Maxwell point, where constraints and degrees of freedom have exactly the same number. One such example is a 2D plate embedded in  the 3D space ($d=2$ and $D=3$). If the bending energy is negligible, such a plate will be called a Maxwell plate.

\subsection{Local embedding and the Janet-Cartan theorem}
\label{sec:janetcartan}
The counting argument discussed above is in full alignment with the Janet-Cartan theorem~\cite{janet1926possibilite,cartan1927possibilite}.

For any given metric $g=g_0$ (which minimizes $E_s$ and thus fully releases built-in stress), whether a mapping $\mathbf{r} \to \mathbf{R}$ exist such that $(g_{0})_{ij}=\partial_i \mathbf{R}\cdot \partial_j \mathbf{R}$, is known as the isometric embedding problem.  $g=g_0$ may not always be achievable, because actual variables of the system are $\mathbf{R}(\mathbf{r})$, and different components of $g$ are related with one another.  
The  Janet-Cartan theorem of analytic local embedding states tells us that a \emph{local embedding} always exists for $D\ge d(d+1)/2$, where local embedding means that for any real space point, we can always find a region around this point such that $g=g_0$ for every point in this region for any analytic $g_0$. In other words, as long as our elastic system is small enough and has open boundaries, any under-constrained or Maxwell systems ($D\ge d(d+1)/2$) must have a stress-free ground state for any analytic $g_0$, i.e., the system has no states of self stress. In contrast, for over-constrained systems $D>d(d+1)/2$, an isometric embedding doesn't exist in general. Physically, this means that for a generic $g_0$, it is impossible to reach $g=g_0$ and fully released stress. Thus, the system contains states of self-stress.  

This is in good agreement with conclusions in the previous section following from Maxwell's counting argument.  It is worth noting that the Janet-Cartan theorem incorporates nonlinearities in the constraint $g=g_0$, and interestingly it supports the linear counting argument.

Below, we consider two examples: an over-constrained system with $d=D$ and a Maxwell plate with $d=2$ and $D=3$, in order to verify relations between zero modes and states of self stress from Eq.~\eqref{eq:N0s}].

\subsection{Over-constrained systems with $d=D$}
According to the counting argument outlined above, a system with $d=D>1$ is always over-constrained, and thus we expect extensive number of states of self stress. 
As shown above, this conclusion qualitatively agrees with the Janet-Cartan theorem.
In this section, we further quantitatively verify the counting argument by directly counting the number of states of self stress and then comparing 
it with the value predicted by the counting argument [Eq.~\eqref{eq:N0s}]. Same as in discrete systems, we start from a fine-tuned stress-free ground state 
(with $g=g_0$) and ask how many different ways residual stress can arise in such a system, which is a linear elasticity problem and thus a definite answer can be obtained.

Because the number of floppy modes per real space point is in general zero without fine tuning in an over-constrained system, 
according to  Eq.~\eqref{eq:N0s}, the number of states of self stress should equal to the number of constraints minus the number of degrees of freedom, 
which is $D(D-1)/2$ for $d=D$. This is indeed correct. A $D$-dimensional stress tensor field $\sigma$  has $D(D+1)/2$ independent components. 
At the same time, force equilibrium condition requires the stress field to be divergenceless,
\begin{align}
 \partial_i \sigma_{ij}=0,
\end{align}
which provides $D$ partial differential equations and thus fixes $D$ of the $D(D+1)/2$ independent components at each real space point. As a result, the number of free components becomes $D(D-1)/2$, matching exactly the expected number of states of self stress of the counting argument. It is worthwhile to mention that to count the states of self stress, in 
principle, force-free boundary conditions also need to be enforced in addition to the divergenceless condition. However, because the boundary is subextensive, 
its impact on an extensive quantity (i.e. the number of states of self stress here) is negligible and thus would not change the counting. 

We conclude this subsection by discussing the sufficient and necessary condition, under which this over-constrained system has a stress free ground state. 
Similar to over-constrained networks/lattices, although generically our over-constrained system shall contain residual stress, the ground state could be made stress free if we fine tune the constraints to make them fully compatible with each other. For lattices and networks this fine tuning is about the rest length of each bond, while in continuum, we need to fine tune the reference metric tensor $g_0$. For $d=D$, the sufficient and necessary condition for an elastic body to be stress free is that $g_0$ must be flat, because the $D$ dimensional Euclidean space we try to embed this body into is flat. More precisely, if we treat $g_0$ as a metric tensor and compute its Gaussian curvature (as show in App.~\ref{app:sec:2Din3D}, for a Riemannian manifold, its Gaussian curvature is uniquely determined by the metric tensor), this Gaussian curvature must vanish at every point, which implies that the system is globally flat. 

\subsection{Maxwell plates and holographic floppy modes}
If $E_b$ is ignored, a 2D plate ($d=2$ and $D=3$) satisfies the Maxwell condition [Eq.~\eqref{eq:z2D}]  with $D=d(d+1)/2$, and thus is right at the Maxwell point, which is the reason
why they are called Maxwell plates.

In this section, we consider small plates with open boundary conditions, where the Janet-Cartan theorem applies. More generic cases, e.g., large or infinite plates and closed plates without boundary, will be discussed in Sec.~\ref{SEC:global}.

As shown above, the number of degrees of freedom and constraints perfectly match in the bulk of a Maxwell plate.  
Thus, edge degrees of freedom become important. Same as Maxwell lattices/networks, a point on an open edge has the same number of 
degrees of freedom but fewer constraints, because its constraints only comes from one side of the edge, while the other side is empty and thus enforces 
no constraint. This deficit of constraints on the boundary leads to floppy modes in Maxwell plates when they are under open boundary conditions, and according to Eq.~\eqref{eq:N0s}, the number of these floppy modes is proportional to the length of the boundary, and is a sub-extensive quantity.
In addition, same as Maxwell lattices, the edge origin implies that floppy modes in Maxwell plates are \emph{holographic}, i.e., for any low-energy deformations, the deformation field at the edge fully dictates the bulk deformation field. The existence of sub-extensive holographic floppy modes is a key property of Maxwell plates, in strong analogy to Maxwell lattices.

These counting-based conclusions on holographic floppy modes can be rigorously proved using  mathematical tools of isometric embedding. As show in Ref.~\onlinecite{Han2006isometric}, 
mathematically, the problem of isometric embedding is translated into partial differential equations, known as the Gauss-Codazzi equations (See App.~\ref{app:sec:2Din3D} for
more details). 
For plates considered here, solutions to these partial differential equations alway exists, as long as suitable boundary conditions are applied and the plate is not too large to hit singularity points. In particular, solutions to these equations have a one-to-one correspondence with the boundary conditions, i.e. boundary conditions fully determine the solution in the bulk, which is the mathematical origin of the holographic floppy modes. This is in strong analogy to static electricity in a cavity, which is also governed by a set of partial differential equations. There, the field configuration is also holographic, i.e. it is fully dictated by boundary conditions, e.g. charge or field configurations on the boundary uniquely determines the charge/field configuration in the bulk.

\subsection{Global embedding and systems beyond the Janet-Cartan theorem}\label{SEC:global}
In this section, we explore infinitely large 2D plates and/or closed 2D plates without open boundaries. Here a global embedding is required to fully release the stress,
which is a much stronger requirement in comparison with a local embedding. A global embedding requires $g=g_0$ even if we extends the system (through analytic continuation) to infinity or until the system forms a closed manifold . A global embedding cannot always be achieved for $D=d(d+1)/2$. 
In the counting argument, this means that at $D=d(d+1)/2$ , although $g=g_0$ introduces no redundant constraints for any small regions, if the elastic system is large enough, or if the system form a closed manifold (e.g. by enforcing periodic boundary conditions), redundant constraints may arise and thus the ground state may not be able to fully release stress unless we fine tune these constraints to make them compatible with one another. 

In this section, we consider two different families of Maxwell plates with (1) positive or (2) negative Gaussian curvature. Remarkably, for both cases, their elastic properties obey the counting predictions.

First, we consider a 2D closed plate with positive Gaussian curvature and zero genus, which means that the plate is topologically equivalent to a sphere $S^2$. Because such a plate has no boundary, the number of constraints and degrees of freedom perfectly match with each other at every point, and thus the counting argument requires that the density of floppy modes and states of self stress must coincide.  
This is indeed true for such a plate. Mathematically, the embedding of such a plate 
(positive Gaussian curvature and zero genus) is known as the Weyl problem~\cite{Weyl1916uber}. Based on the Nirenberg-Pogorelov theorem~\cite{Nirenberg1953,Pogorelov1952}
such an embedding with $g=g_0$ always exists and the embedding is ``rigid'',  i.e., any deformation from the ground state will make $g\ne g_0$ and thus cost energy.
The existence of embedding for any $g_0$ implies that the ground state is always stress free, and thus there is no state of self stress $N_s=0$. At the same time, the rigidity
part of the Nirenberg-Pogorelov theorem implies the absence of floppy mode ($N_f=0$), and thus indeed $N_f=N_s$ as the counting argument predicts.
In addition, because such a plate has $N_f=N_s=0$, this system is not only ``Maxwell'', but also ``isostatic''.

For plates with negative Gaussian curvature, it cannot form a closed manifold and floppy modes can emerge from edge degrees of freedom as shown in the section above. Here, we consider an infinitely-large plate. If the Gaussian curvature decays to zero fast enough at infinity, a global embedding always exits~\cite{Hong1993} and thus 
$g=g_0$ can always be reached. Here, conclusions of local embedding remain and we shall have no state of self-stress and sub-extensive number of floppy modes. 
If the Gaussian curvature doesn't approach zero or approaches zero too slowly at infinity, an isometric embedding may hit singularity and thus make a global embedding
impossible (e.g. the Efimov theorem~\cite{Efimov1966surfaces}). Because $g$ cannot reach $g_0$, the ground state carriers residual stress, i.e., state of self stress emerges.
As shown in Ref.~\cite{marder2006geometry,sharon2010mechanics},  wrinkles will start to develop beyond these singular points. According to Maxwell's counting argument, this emergence of state of self-stress must be accompanied by new floppy modes. For this example, these floppy modes correspond to translations of the wrinkles.

\section{2D plates with bending energy}\label{SEC:PlatBend}
In this section, we consider the impact of bending energy $E_b$ for $d<D$. As mentioned above, we will focus on the thin plate limit (small $h$) and thus $E_b \ll E_s$.
In the counting analysis, $E_b$ can also be treated as constraints. 
Same as $E_s$, generally, the number of constraints enforced by $E_b$ is an extensive quantity. For 2D plates in a 3D space, as shown in the previous section, without $E_b$, the number of degrees of freedom exceeds the number of constraints by a small (sub-extensive) margin, and as a result, after adding extensive number of constraints from $E_b$, the system is over-constrained. Therefore, same as the case of $d=D$, we shall in general expect one unique ground state (no floppy mode) and extensive number of states of self-stress.  Thus, unless we fine tune $g_0$, the system shall in general carry residual stress.

In this section, we answer two questions:
\begin{itemize}
\item Is it possible to make a 2D plate with bending stiffness stress free? 
\item Is it possible to make a 2D plate with bending stiffness flexible, i.e. introduce one or more floppy mode(s) into this over-constrained system? 
\end{itemize}
It turns out that the answer to the second question relies on the first one, and thus we will start by exploring the first question.

\begin{figure}[t]
	\centering
	\subfigure[]{\includegraphics[width=.4\columnwidth]{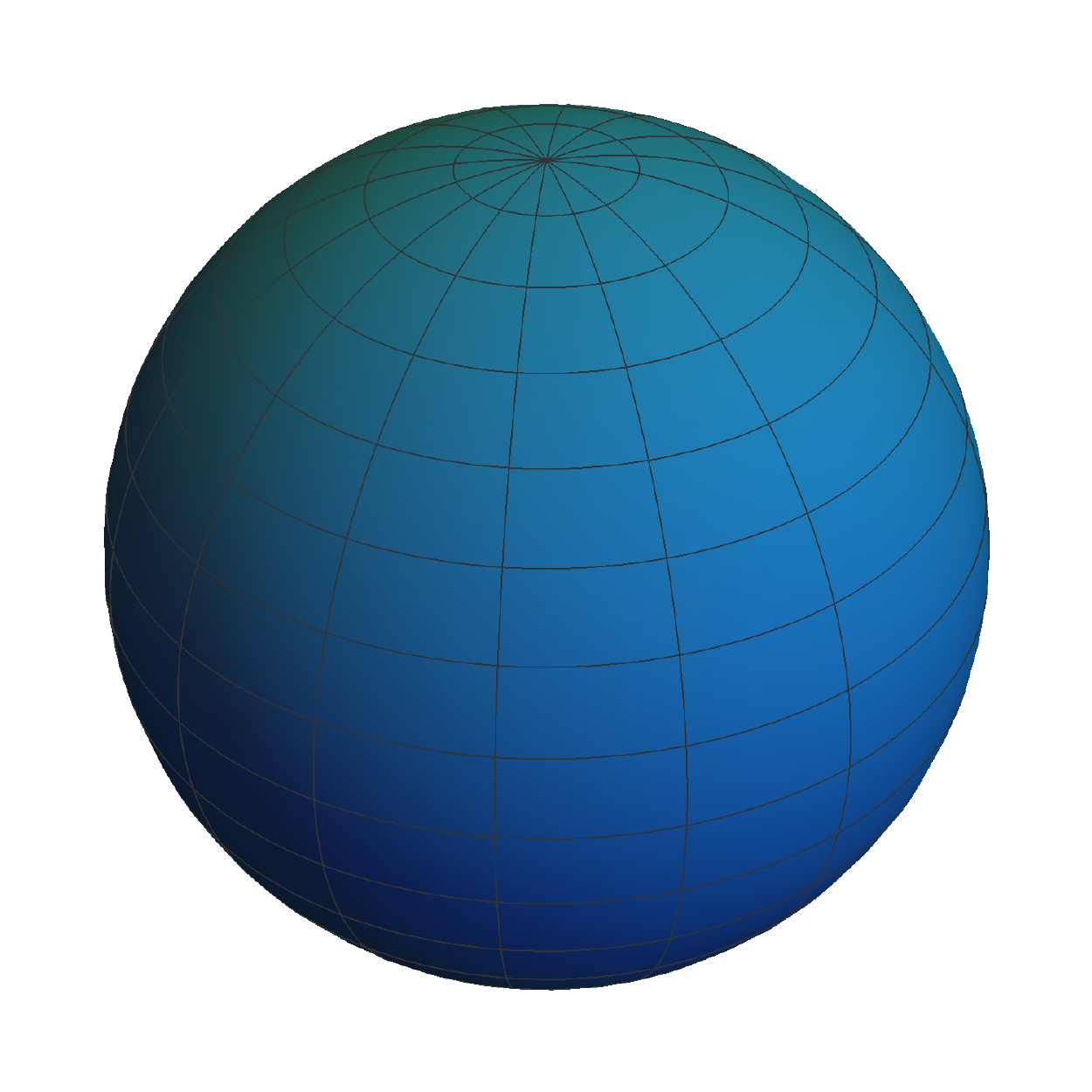}}
	\subfigure[]{\includegraphics[width=.4\columnwidth]{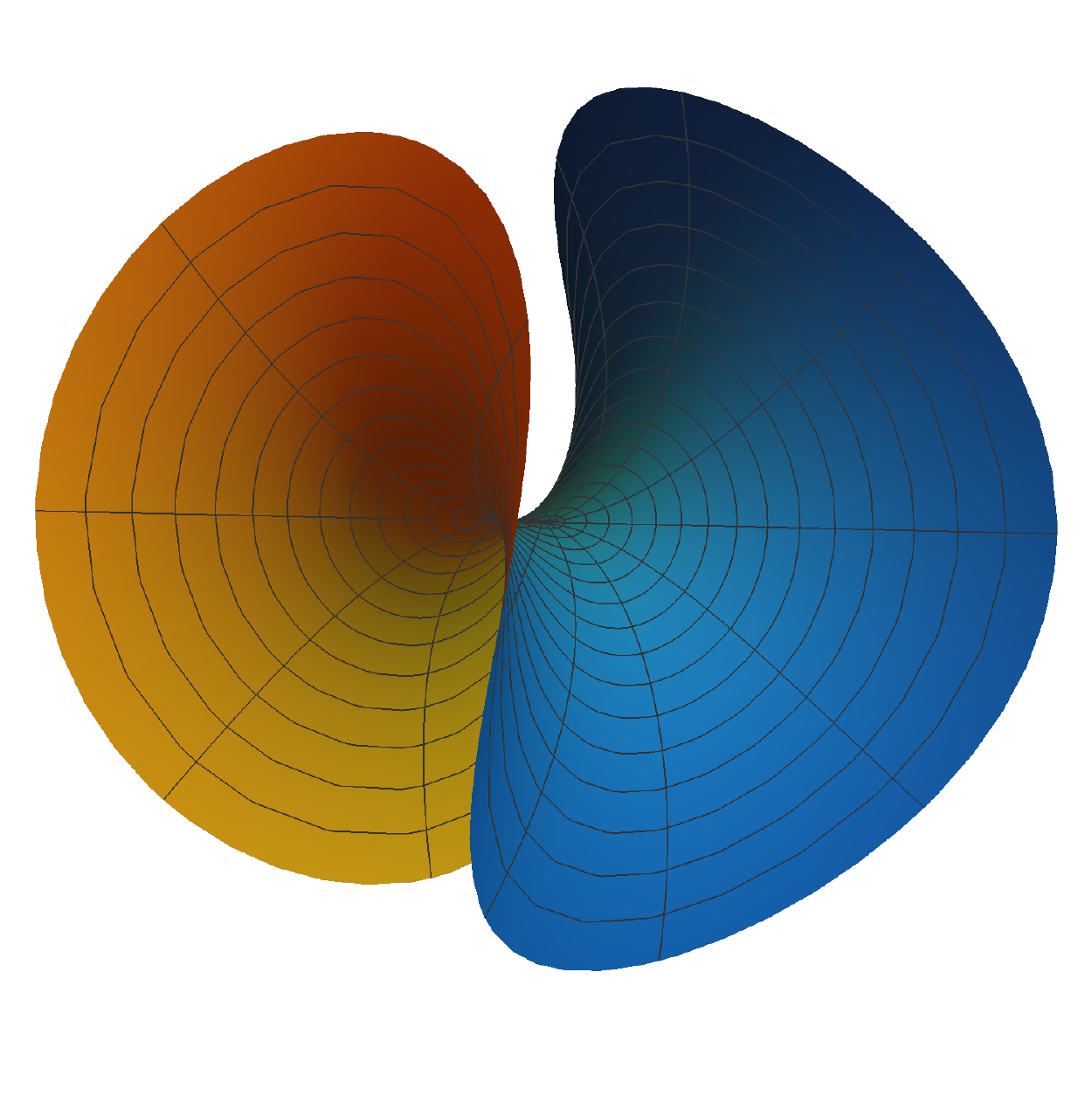}}
	\caption{Two types of stress-free 2D plates with (a) positive or (b) negative Gaussian curvature. (a) With a positive Gaussian curvature, a sphere is the only configuration for 2D stress-free plates. (b) With a negative Gaussian curvature, a 2D plate must form a minimal surface to be stress-free. Here, we present one such example, an Enneper surface.}
	\label{fig:}
\end{figure}

Here we show that a 2D plate becomes stress free, if and only if $g_0$ satisfies either one of the following two criteria:
\begin{itemize}
\item the sphere criterion: the Gaussian curvature $K_0$ calculated from the metric tensor $g_0$ is a non-negative constant over the entire plate.
\item the minimal surface criterion: if $K_0<0$, we treat the product $\sqrt{|K_0|}g_0$ as a new metric tensor and compute its Gaussian curvature $K_1$. $K_1$ must vanish at every point on the plate.
\end{itemize}
If the first criterion is met, the ground state of such a 2D plate is (part of) a sphere with radius $\rho_0=1/\sqrt{K_0}$. Here, the system is stress free and it has a unique ground state (no floppy mode). 
As proved by Ricci in 1894, the second criterion is the sufficient and necessary condition for the existence of a minimal-surface with $g=g_0$~\cite{Ricci1894}. 
If $g_0$ satisfies the second criterion, the ground state is a minimal surface. Remarkably, such a 2D plate is not only stress free but also has one floppy mode [see Sec.~\ref{SEC:MiniSurf} below]. One interesting example of minimal-surface plates 
can be found in Ref.~\onlinecite{Efrati2011}, which applies to any narrow ribbon with negative Gaussian curvature.
If none of these two criteria holds, the corresponding 2D plate must have residual stress, which cannot be released.

Here, we just provide a brief outline of the proof, while details can be found in App.~\ref{app:sec:stress:free:2D}. Within the perturbative approach described above (keeping terms up to $O(h^3)$), in order to find the ground state, we first minimize the leading-order term $E_s$, which enforces the constraint $g=g_0$, and then, under the constraint of $g=g_0$, $E_b$ is then minimized. The constraint $g=g_0$ not only fixes $g$ but also the Gaussian curvature $K$ of the 2D plate setting $K=K_0$, because the Gaussian curvature  is uniquely determined by the metric tensor and its spatial derivatives (known as Gauss’s  \emph{theorema egregium}, see App.~\ref{app:sec:2Din3D} for more details). For $E_s$, this means that the value of $K$ in Eq.~\eqref{eq:Eb} is already pinned as we set $g=g_0$. As a result, the $K$ term in $E_b$ does not participle in the minimization of $E_b$ and thus can be ignored. After dropping $K$,  it is straightforward that $E_s$ prefers to minimize the square of the mean curvature $H=(k_1+k_2)/2$, where $k_1$ and $k_2$ are the two principle curvatures. This minimization needs to obey the constraint on the Gaussian curvature $K=k_1 \times k_2=K_0$. With a fixed $k_1\times k_2$, the minimization of $H^2$ has two possible scenarios: (1) $k_1=k_2=\sqrt{\kappa_0}$ for $K \ge 0$ or (2) $k_1=-k_2$ for $K<0$. This is the constraint enforced by $E_b$.

As shown in App.~\ref{app:sec:stress:free:2D}, $E_s$ enforces constraints on the metric tensor, which is also known as the first fundamental form, while constraints enforced 
by $E_b$ are about curvatures, which are mainly associated with the second fundamental form. Thus, in order to ensure constraints enforced by $E_s$ and $E_b$ are 
compatible with each other, relations between first and second fundamental forms play a crucial role. It is known that the first and second fundamental forms are related by a set of partial differential equations, known the Gauss-Codazzi equations (in addition to the constraint of Gaussian curvature discussed above). As shown in App.~\ref{app:sec:stress:free:2D}, the Gauss-Codazzi equations are satisfied if and only if one of the two criteria mentioned above is met.

\section{Minimal surfaces and their floppy mode}\label{SEC:MiniSurf}
As shown in the previous section, when bending energy is taken into account, 2D plates are generically over-constrained and only two special cases can be stress free (a sphere or a minimum surface).  Elastic properties of these two cases are very different. Most importantly, a sphere has a unique ground state, while a minimal-surface 2D plate has infinitely many degenerate ground states, which are connected with each other via a floppy mode. In this section, we focus on the case of minimal surfaces.

\subsection{Floppy mode}
Minimal surfaces are 2D surfaces embedded in a 3D space with their local area minimized. The key signature of a minimal surface is the vanishing mean curvature $H=0$. For $E_b$, because the Gaussian curvature $K$ is fixed by $g_0$ as we enforce the constraint $g=g_0$, a minimal surface with $H=0$ offers a natural minimization of $E_b$. This is the key reason why minimal surface can make constraints from $E_b$ and $E_s$ compatible. 

As shown in App.~\ref{app:sec:minimal:surface}, for any minimal surface, an isometric deformation always exists (i.e. $g$ stays invariant), under which the surface remains minimal (i.e. $H$ remains $0$). One example of such an isometric deformation is shown in Fig.~\ref{fig:associate:family}.
For elasticity, it is easy to notice that this isometric deformation doesn't cost any elastic energy, as $E=E_s+E_b$ stays the same, and thus is a floppy mode. All minimal surfaces that are connected by this deformation is called an associate family, and this floppy mode transfers one minimal surface to other minimal surfaces, which belong to the same associated family.

Minimal surfaces in an associate family can be labeled by a phase angle $\varphi$ (see App.~\ref{app:sec:minimal:surface} for details). As a phase angle, only $\varphi \mod 2\pi$ has a real physical meaning. 
In the  Weierstrass-Enneper parameterization, minimal surfaces are mapped into complex functions, and the phase angle $\varphi$ is an overall complex phase in the Weierstrass-Enneper parameterization. By varying this phase angle (and keeping the rest of the complex functions invariant), a minimal surface is deformed into other ones in its associate family.
In this parameterization, the floppy mode corresponds to adiabatically adjusting the value of $\varphi$.

It is worthwhile to emphasize that although for a minimal surface 2D plate $E=E_s+E_b$ has infinite degenerate ground states, in reality, this infinite degeneracy will be lifted by higher order terms in the elastic energy (e.g. if one introduces anisotropy into the elastic medium and one example is shown in App.~\ref{app:sec:higher:order}). 
As a result, a real system may prefer one (or two) specific minimal surface(s) as its ground state, while the floppy mode becomes a soft mode, i.e., it costs a small amount of energy to deform the ground state to another minimal surface in its associate family. However, as long as we are focusing on low-energy soft deformation, other configurations beyond the associate family can still be excluded, which cost much higher energies to reach.

\subsection{Narrow ribbons}\label{SEC:Ribb}
\begin{figure}[t]
	\centering
	\includegraphics[width=.9\columnwidth]{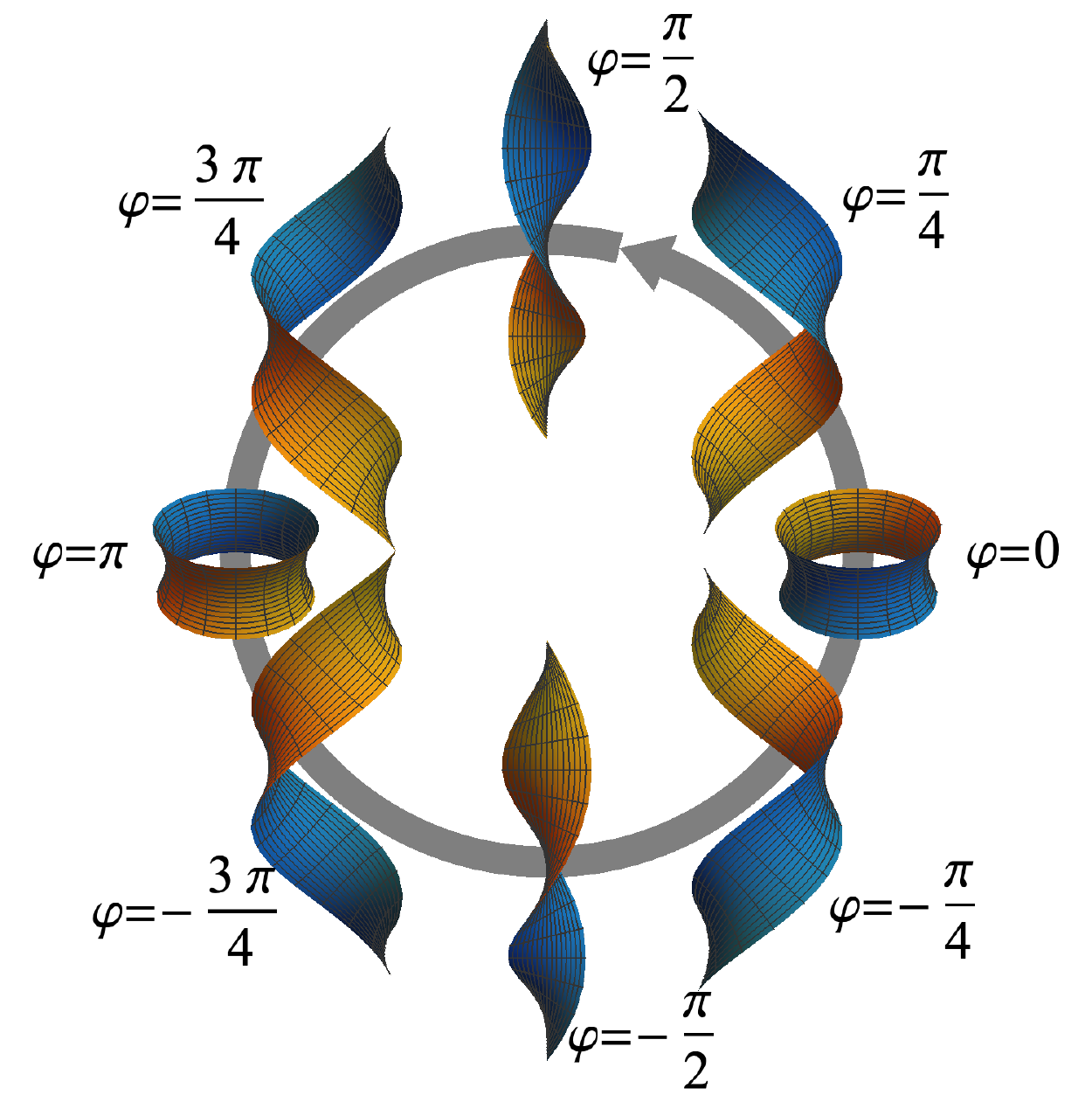}
	\caption{Ground state configurations of 2D narrow ribbon with negative Gaussian curvature. Any minimal surface in the helicoid-catenoid associate family is a ground state. These ground states
	are labeled by a phase angle $\varphi$ as marked in the figure.}
	\label{fig:associate:family}
\end{figure}

Here, we present one example of stress-free 2D plates, by considering one easy realization of the stress-free criteria. As shown in App.~\ref{app:sec:ribbons}, for  2D ribbons, as long as the width of a ribbon is narrow enough, the system is guaranteed to satisfy one of the two stress-free criteria regardless of microscopic details, and thus their ground states must be either part of a sphere or a minimal surface. Interesting examples of of minimal surface hyperbolic thin ribbons have been discussed in Ref.~\onlinecite{Efrati2011}.

In this section, we focus on 2D narrow ribbons with minimal surface ground states.
As discussed above, such a system has infinite number of ground states characterized by a phase angle $\varphi$. As shown in in Fig.~\ref{fig:associate:family}, 
$\varphi=\pm \pi/2$ represent the helicoid ground states with left-hand and right-hand chirality respectively, while $\varphi=0$ and $\pi$ represent two catenoid ground states. Other values of $\varphi$ represent other ground states in this minimal-surface associate family, known as the helicoid-catenoid family.
It is worthwhile to mention that as shown in App.~\ref{app:sec:ribbons}, depending on the linear term of $u$ in $g_{22}$ [Eq.~\eqref{app:g22_expansion}],
a ribbon's helicoid ground state may or may not contain the rotational axis, i.e. helicoids shown in Fig.~\ref{fig:associate:family} at $\varphi=\pm \pi/2$ vs. helicoids in Fig.~\ref{fig:spin:chain}(e) and (f). These two cases may look different, but they are part of the same helicoid and thus share the same elastic properties.  Specifically, helicoids in Fig.~\ref{fig:spin:chain}(e) and (f) can be seen as cutting a thin strip from the edge of helicoids in Fig.~\ref{fig:associate:family}.

As can be seen from the figure, except for the two catenoid ground states ($\varphi = 0$ or $\pi$), all other ground states of the 2D plate simultaneously break the chiral symmetry (of the target space). 
The two catenoid ground states don't break spatial symmetries of the target space. Instead, they break an internal symmetry of the 2D plate. This can be made more visible if we color the two sides of the plate with two different colors [yellow and blue in Fig.~\ref{fig:associate:family}]. In our elastic energy, the two sides are fully equivalent and thus the elastic energy is invariant if we swap these two colors. However, a catenoid ground state breaks this symmetry making the yellow and blue inequivalent. Because swapping the two colors is a $Z_2$ symmetry, we should expect two such ground states when this symmetry is broken, which are the two catenoids at $\varphi= 0$ and $\pi$. If we swap the two colors, the two catenoids are mapped into each other. This is in analogy to an $Z_2$ internal symmetry breaking, e.g. a ferromagnetic state in the Ising spin model.

\subsection{Fractional solitons in thin ribbons}
The floppy mode in 2D minimal-surface plates has one interesting implication. It means that the system supports fractional topological excitations, in strong analogy to 
 fractional topological states such as a $Z_2$ spin liquid.

Because the system has infinite degenerate ground states $0 \le \varphi<2\pi$, low-energy excitations in such a ribbon shall be characterized as a $\varphi$ field slowly varying   along the ribbon.
The elastic energy for such a low-energy excitation can be described by the following phenomenological theory
\begin{align}
E=\int d v (\partial_v \varphi)^2
\end{align}
where $v$ is the coordinate along the ribbon direction.
Because any constant $\varphi$ produces a degenerate ground state, this elastic energy can only depend on derivatives of $\varphi$, instead of $\varphi$ itself.  The quadratic term included here is the leading order contribution. Without loss of generality, the elastic modulus for this term is set to unity, which describes the energy cost for an inhomogeneous $\varphi$.

As mentioned above, in a real system, the infinite number of minimal-surface states in the associate family cannot be perfectly degenerate, due to higher order corrections in the elastic energy, which can be described by a function of $\varphi$, $E_{c}(\varphi)$. Because for a phase angle, $\varphi$ and $\varphi+2\pi$ are fully equivalent as mentioned above, this function must be a periodic function with $E_c(\varphi)=E_c(\varphi+2\pi)$. Thus, in general, the elastic energy shall take the following form
\begin{align}
E=\int d v \left[ (\partial_v \varphi)^2+ \sum_n \gamma_n \cos  n (\varphi-\varphi_{n})\right]
\label{eq:sine_gordon_general}
\end{align}
with the second term describing higher order terms that lift the infinite degeneracy of the ground state. Here, $n$ are positive integers, while $\gamma_n$ and $\varphi_{n}$ are constants for each $n$. This effective theory is a generalized sine-Gordon theory. It has long been known that the sine-Gordon theory supports soliton excitations, i.e., topological excitations 
with $\varphi$ changes from one value to another over a finite length scale. A particularly interesting example of a soliton in a mechanical chain described by a sine-Gordon equation is the nonlinear soliton in a 1D topological rotor chain, as discussed in Ref.~\onlinecite{Chen2014}, where the kink soliton is exactly zero energy due to a boundary term.  

It must be emphasized that the sine-Gordon theory we use here  is \emph{compact}, i.e. $\varphi$ and $\varphi+2\pi$ are identical. As will be discussed below, solitons are quantized in a compact sine-Gordon theory, and each soliton has a well-defined quantized charge. 
Here, we first introduce the definition of soliton charge, while its quantization will be addressed below. For a compact field $\varphi$, the periodicity $2\pi$ offers a natural reference to measure the varition of $\varphi$. For a soliton where $\varphi$ changes by $\Delta \varphi$, its charge is defined as $\Delta \varphi /2\pi$. Solitons with integer charge are called integer solitons, i.e., $\Delta \varphi = 2 n \pi$ with $n$ being an integer, while a fractionally charged  soliton will be called a fractional soliton.

For a (generalized) sine-Gordon theory shown in Eq.~\eqref{eq:sine_gordon_general}, generically, solitons  have integer charge (i.e., $\varphi$ changes by $2 n \pi$). 
This charge quantization is because generically these cosine terms will select a unique ground state with a specific value of $\varphi$ (say $\varphi=\varphi_0$). Thus, for a soliton, in order to ensure that the system is at the ground state away from the soliton, $\varphi$ must change by $2\pi$ times an integer (e.g. from $\varphi_0$ to $\varphi_0+ 2 n \pi$). This leads to the quantization of the soliton charge. Here, we demonstrate one such example by keeping only the $n=1$ term in the elastic energy, where the standard sine-Gordon theory is recovered $E=\int d v \left[ (\partial_v \varphi)^2+\gamma \cos \varphi\right]$. Such a theory is known to support soliton excitations with $\Delta \varphi =2\pi$, i.e. charge-$1$ solitons. 
By combining multiple charge-1 solitons together, solitons with higher integer charge can be created, but solitons with non-integer (fractional or irrational) charge are prohibited.

For 2D plates that we considered here, however, the situation differs sharply from the generic case discussed above. 
As shown in App.~\ref{app:sec:minimal:surface}, the transformation $\varphi \to \varphi+\pi$ is a chiral symmetry transformation ($\mathbf{R} \to -\mathbf{R}$). 
Because the elastic energy of a 2D plate is invariant under a chiral transformation, our elastic energy must be invariant under $\varphi \to \varphi+\pi$ (although the ground states can spontaneously break this symmetry, as we discuss in Sec.~\ref{SEC:Ribb}), and thus 
$E_c(\varphi)=E_c(\varphi+\pi)$, in contrast to a generic compact sine-Gordon theory where $E_c(\varphi)=E_c(\varphi+2\pi)$. This means that with the help of the 
chiral symmetry, our system always contains two degenerate ground states (with $\varphi=\varphi_0$ and $\varphi=\varphi_0+\pi$), and thus one can introduce
half-charged solitons, which is the domain wall between these two ground states. Because the values of $\varphi$ for these two ground states differ by $\pi$, such a soliton 
must have $\Delta \varphi =(2n+1)\pi$, and thus a fractional charge $n+1/2$. This fractionalization is in analogy to nematic liquid crystals, where the molecules (and the order parameter)
are invariant under a $\pi$ rotation. Such a symmetry also results in fractional topological defects in nematic liquid crystals, i.e., disinclinations or disinclination lines, which can be viewed as
half of a vortex or a vortex line~\cite{degennes1993, Lubensky2000}.

The configuration of these fractional solitons can be obtained using the elastic energy. Because the chiral symmetry requires $E_c(\varphi)=E_c(\varphi+\pi)$, 
 odd-integer cosine terms in Eq.~\eqref{eq:sine_gordon_general} are prohibited and thus we have
\begin{align}
E=\int d v \left[ (\partial_v \varphi)^2+ \sum_n \gamma_{2n} \cos  2 n (\varphi-\varphi_{2n})\right] .
\label{eq:sine_gordon_general_chiral}
\end{align}
Here, we demonstrate a fractional soliton by considering the minimal model, which only contains the $\cos  2 \varphi$ term
\begin{align}
E=\int d v \left[ (\partial_v \varphi)^2+\gamma \cos 2 \varphi\right] .
\label{eq:sine_gordon}
\end{align}
This is similar to  the standard sine-Gordon theory, but with an extra factor of $2$ in the cosine term.  For $\gamma>0$, the system has two ground states $\varphi=\pm \pi/2$,  while for $\gamma<0$, the two ground states are $\varphi=0$ or $\pi$. Here, we demonstrate the physics by focusing on the case of $\gamma>0$. 
Following the standard sine-Gordon theory approach,  soliton solutions are expected. Because of the extra factor of $2$ in the cosine term, $\varphi$ changes by $\pi$ for such a soliton, instead of $2\pi$, and thus it contains half charge. Here, the soliton is the domain boundary between the left- and right- handed helicoid ground states. 

This physic picture of fractional excitations can be generalized.  As mention above, for a system whose low-energy 
physics is described by a compact sine-Gordon theory,  generically, soliton charge is quantized to integer values. However,
if the Hamiltonian is invariant under the $Z_N$ transformation $\varphi\to \varphi +2\pi/N$  ($N$ is an integer),
this $Z_N$ symmetry will change the charge quantization from integer to fractional values (integer times $1/N$). A minimal-surface 2D plate offers such an example with $N=2$, and the $Z_N$ symmetry here is the $Z_2$ 
chiral symmetry $\varphi\to \varphi +\pi$.

\subsection{Numerical verification}\label{SEC:Nume}

\begin{figure}[t]
	\centering
	\subfigure[]{\includegraphics[width=.2\columnwidth]{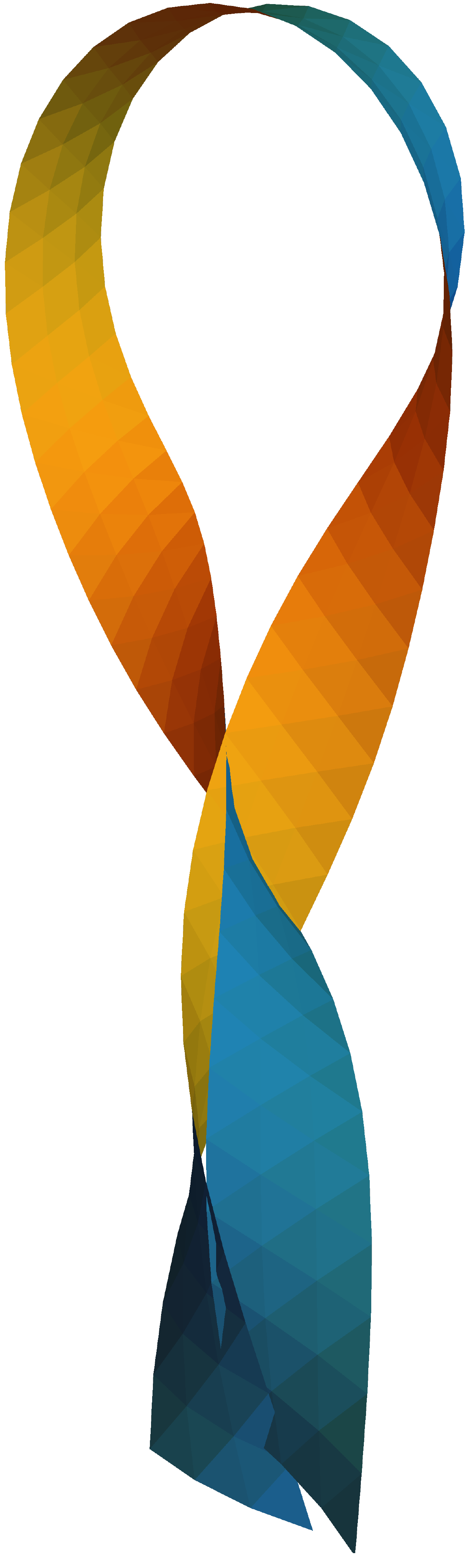}}
	\subfigure[]{\includegraphics[width=.5\columnwidth]{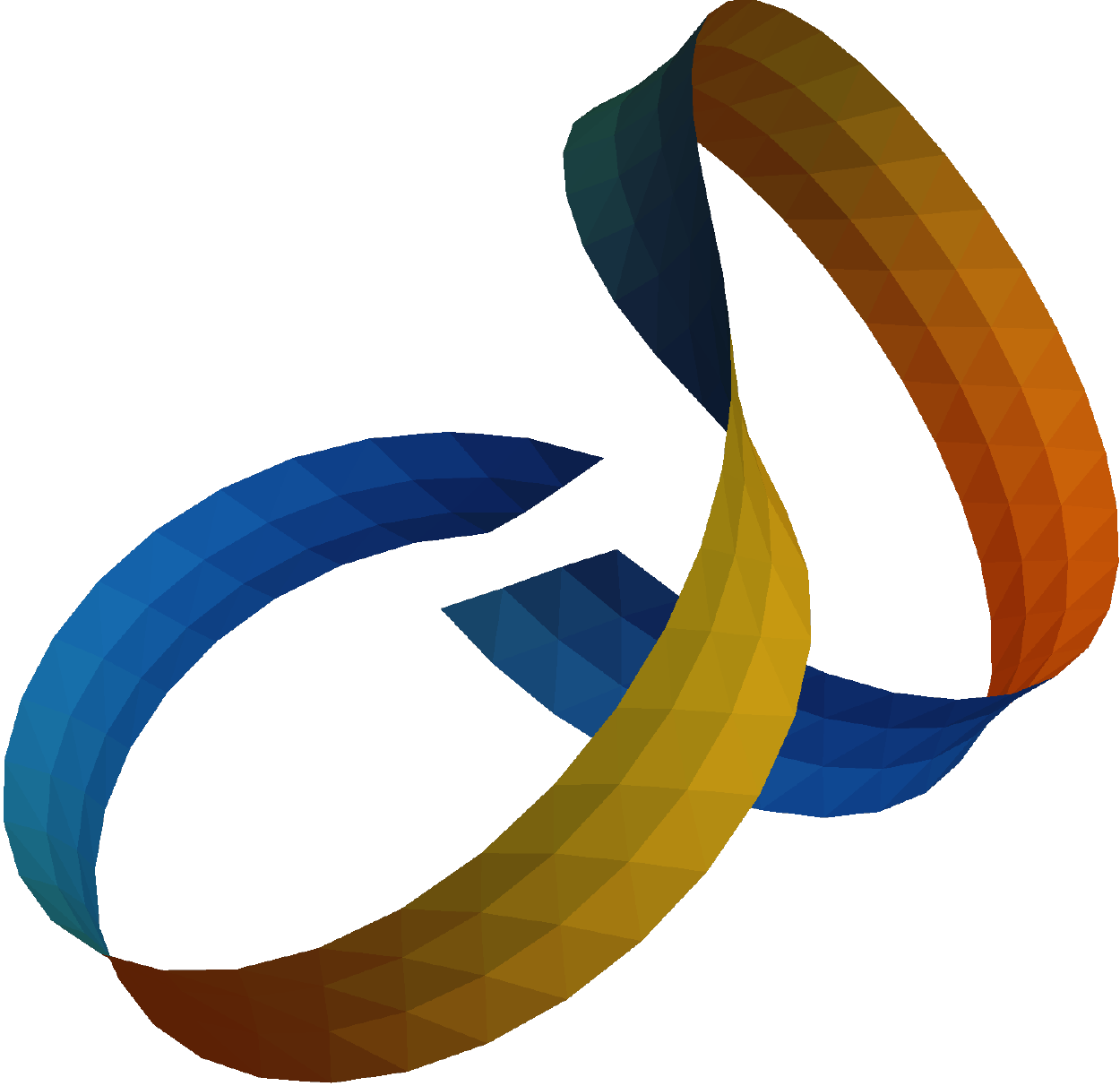}}
	\caption{Soliton configuration from finite-element analysis. By minimizing the elastic energy, here we show the soliton configuration for ribbons with (a)  helicoid or (b) catenoid ground states.}
	\label{fig:numerical}
\end{figure}

These fractional solitons indeed exist in narrow elastic ribbons. 
In Fig.~\ref{fig:numerical}, we simulate a narrow ribbon with $E=E_s + E_b$ and a small perturbation is added to lift the infinite degeneracy of the ground states.
In this simulation, 10-node triangular elements are utilized. The shape function of such an element preserves the three-fold rotational symmetry, which help minimizing 
anisotropy induced by the shape function. The entire ribbon is composed of $60\times 4$ nodes. The elastic moduli are set to (in arbitrary units) 
$h B_0 = 6 \times 10^5$ and $G_0 =B_0/2$  [Eq.~\eqref{eq:Es_2D}]. The bending stiffness 
$D_1=\frac{2 h^3 G (3B+G)}{3(3B+4G)}$ is set to $ 2.4 \times 10^3$ [Eqs.~\eqref{eq:Eb} and~\eqref{eq:Eb:app}]. 
We also added a small perturbation to favor the helicoid (or catenoid) ground states [as shown in Eq.~\eqref{eq:deltaEb:app}], 
whose coefficient $\delta D=0.02 D_1$ in Fig.~\ref{fig:numerical}.(a) and $-0.01D_1$ in Fig.~\ref{fig:numerical}(b).
As shown in App.~\ref{app:sec:ribbons}, $g_0$ can always be written in the form of 
Eqs.~\eqref{app:g0_general} and~\eqref{app:g22_expansion} and here we choose $a_1=0$ and $a_2= 2\pi/10$ [Eq.~\eqref{app:g22_expansion}].
All qualitative features that we observed are insensitive to microscopic details and remain stable as we vary the control parameters and the system size.
The simulation didn't enforce the excluded-volume condition, and thus the ribbon may intersect with itself. 
Enforcing excluded volume or not doesn't change any qualitative conclusions.

From the finite-element analysis, we found that a fractional soliton is indeed a local energy minimum. 
In particular, for the helicoid ground states, 
by minimizing the elastic energy, we find that such a domain structure
always bends the ribbon by nearly $180^\circ$, i.e. if we move along the direction of the helicoid ribbon, each soliton excitation implies a sharp U-turn. The reason for this sharp turn can be understood by looking at a section of catenoid and trying to connect it with two helicoids with opposite chirality to its two ends. As shown in Fig.~\ref{fig:associate:family}, if one tries to smoothly connect them (by slowly varying $\varphi$ along the chain), a $180^\circ$ degree turn shall naturally arise. 
This sharp U-turn associated with such a fractional soliton is the key for the formation of kinks in tangled phone-cords.

\section{Majumdar-Ghosh model and Z$_2$ topological spin liquid}

\begin{figure}[t]
	\centering
	\includegraphics[width=1\columnwidth]{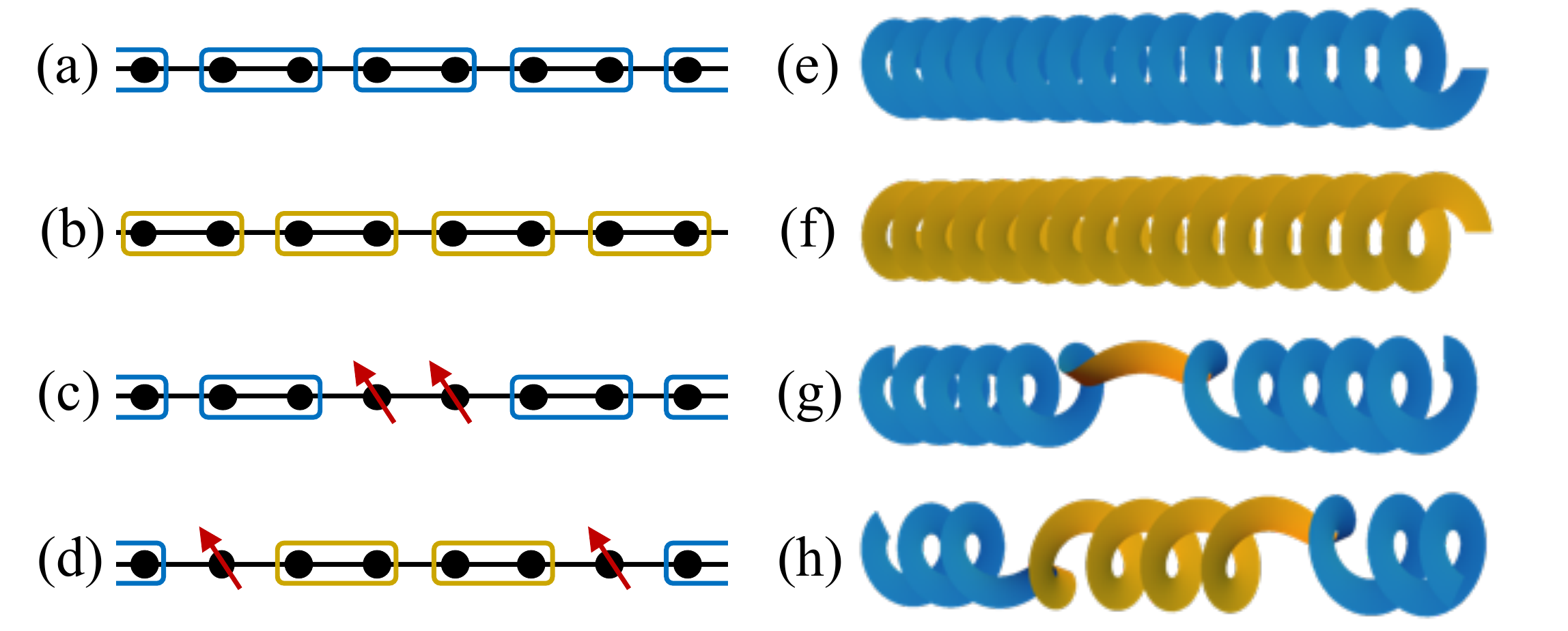}
	\caption{Fractional solitons in dimerized spin chains. (a-d) and narrow ribbons (e-h). The quantum spin system has two degenerate ground states (a) and (b). A spin-1 excitation
	can be created via local perturbations (c), which splits into two deconfined spin-1/2 excitations (d). (e) and (f) show two degenerate ground states of a helicoid ribbon. 
	(g) shows a local excitation, which decays into a pair of fractional solitons in (f).}
	\label{fig:spin:chain}
\end{figure}

It turns out that the elastic system we discuss here and its fractional soliton excitations show strong analogy to fractional excitations in $Z_2$ spin liquids and other similar quantum dimer systems.  Thus we can utilize insights obtained from the study of fractional excitations in quantum systems to help understanding fractional excitations in 2D plates.
In this section, we provide a brief review about some key properties of 2D $Z_2$ spin liquids and 1D quantum dimer systems, which will be compared with 2D plates in the
next section.

A $Z_2$ spin liquid in 2D is one of the most important and well-studied fractional topological states, which exhibit exotic properties such as $Z_2$ topological order,  long-range entanglement, fractional excitations and topological degeneracy (see for example a recent review Ref.~\onlinecite{Wen2017} and references therein).
The study of $Z_2$ spin liquids originates from Anderson's resonating-valence-bond (RVB) senario~\cite{Anderson1973, Fazekas1974} in frustrated quantum spin systems and quantum dimer models~\cite{Kivelson1987, Rokhsar1988, Moessner2001}. 
This exotic quantum phase of matter is characterized by a topological Ising gauge theory and gives rise to deconfined fractional excitations, e.g. spinons which carry spin-$1/2$ 
but no charge~\cite{Read1991,Wen1991,Mudry1994, Senthil2000, Moessner2001b}. Later, an exactly sovable model with the same topological order was introduced, known as the toric code model of Kitaev~\cite{Kitaev2003}.

\begin{figure}[t]
	\subfigure[]{\includegraphics[width=.4\columnwidth]{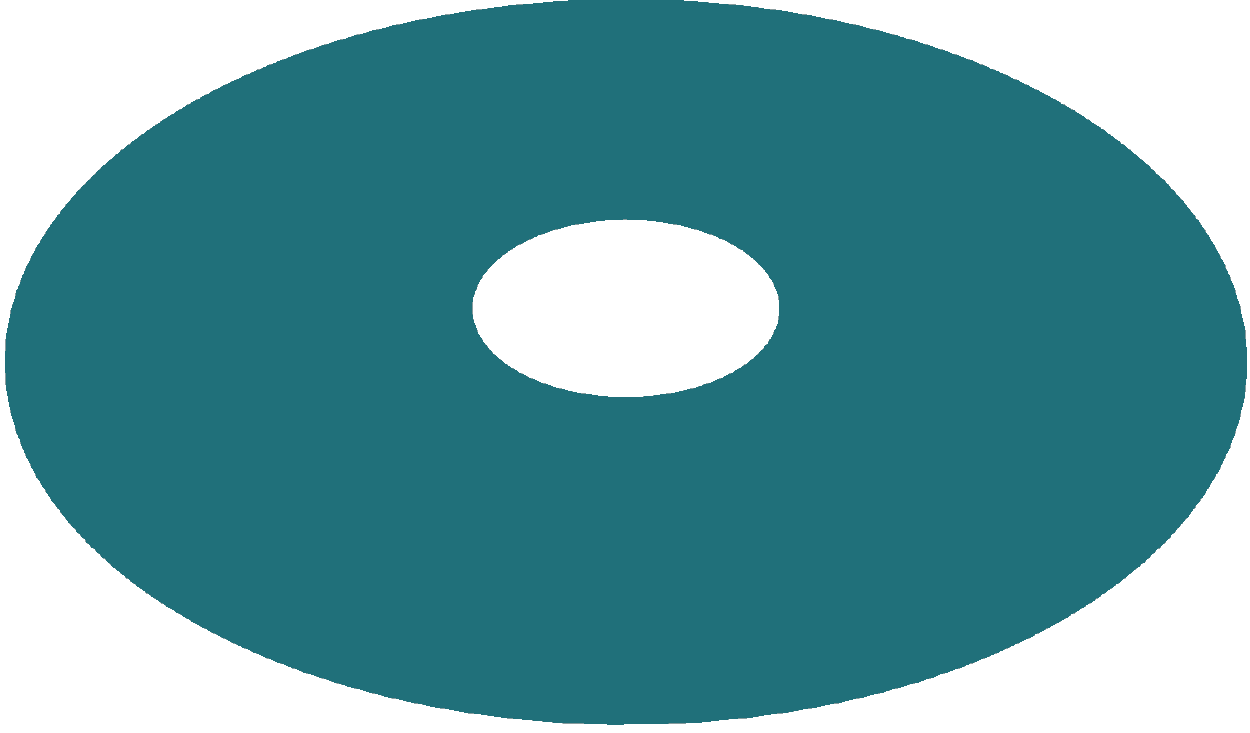}}
	\subfigure[]{\includegraphics[width=.4\columnwidth]{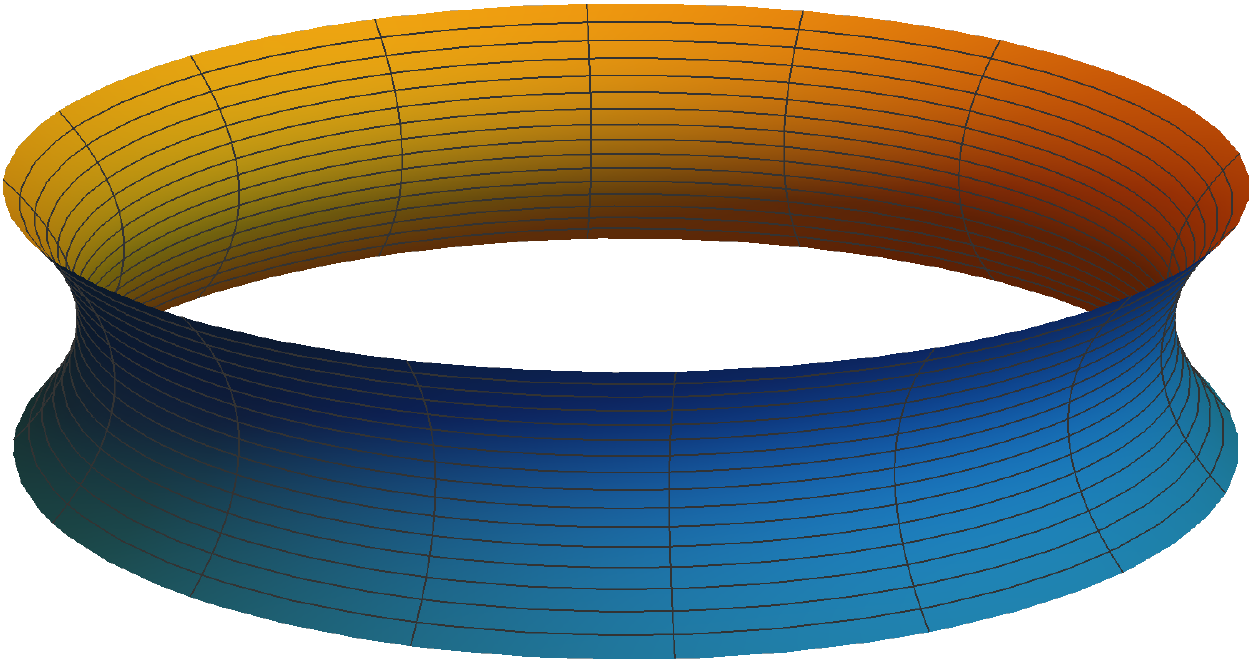}}
	\subfigure[]{\includegraphics[width=.4\columnwidth]{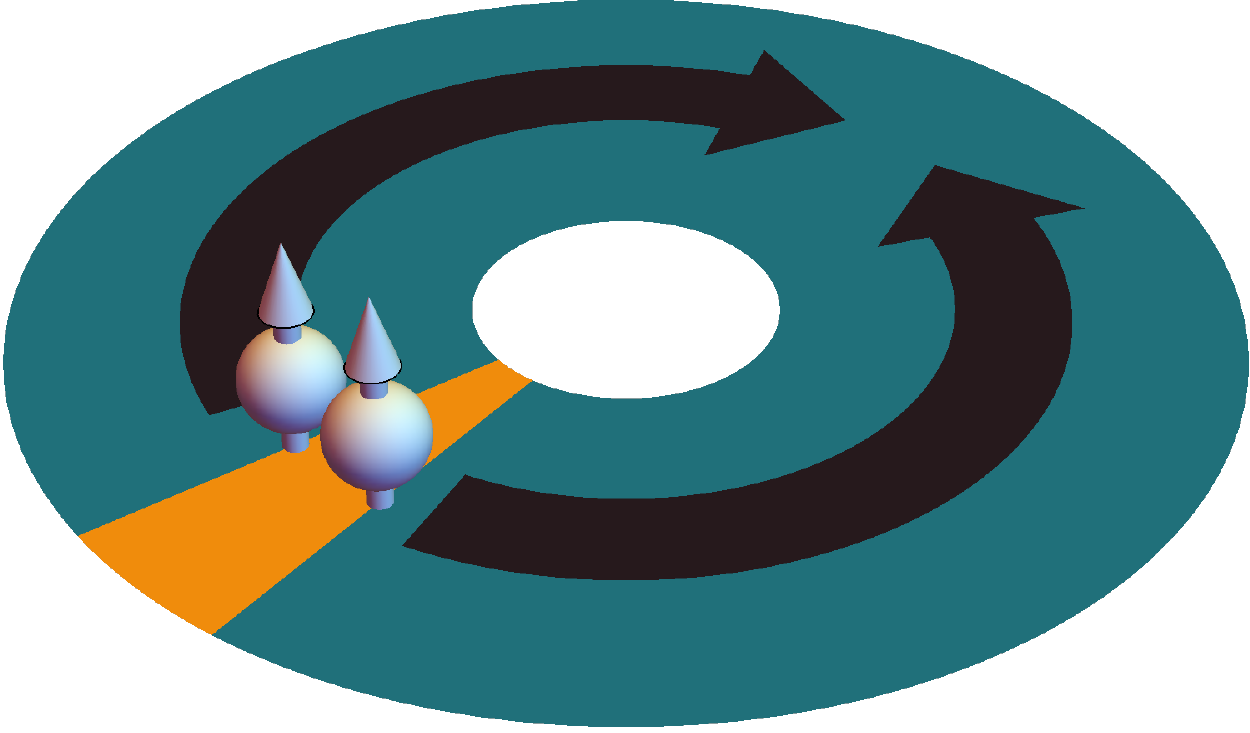}}
	\subfigure[]{\includegraphics[width=.4\columnwidth]{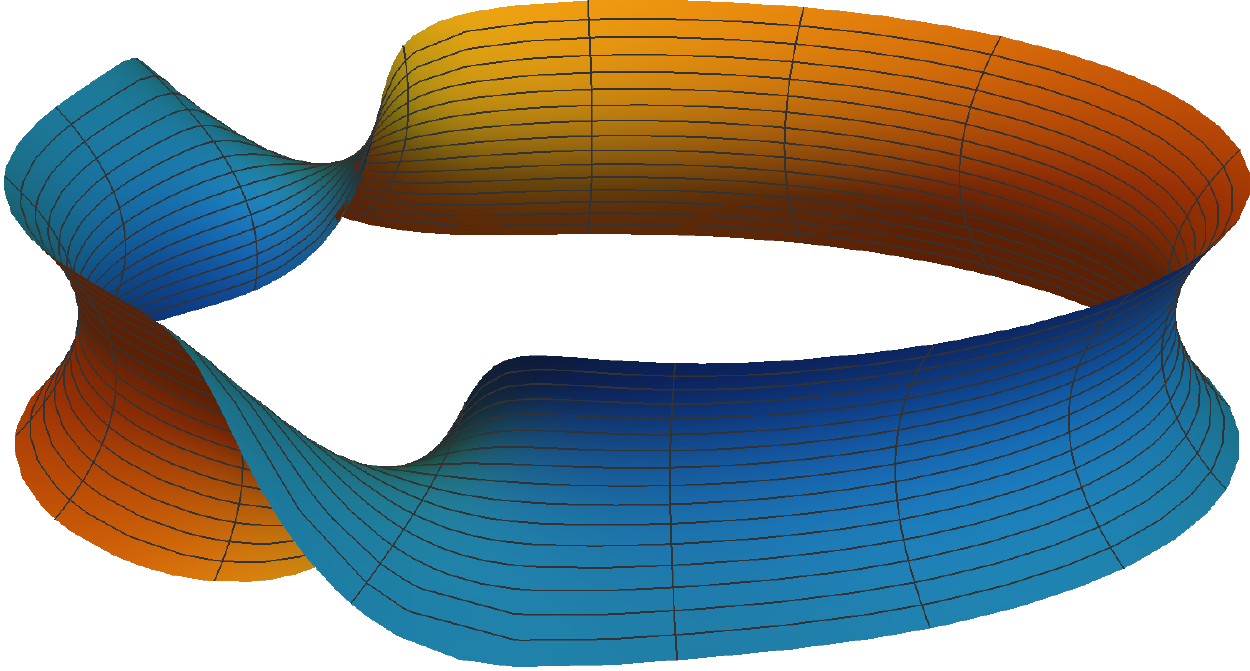}}
	\subfigure[]{\includegraphics[width=.4\columnwidth]{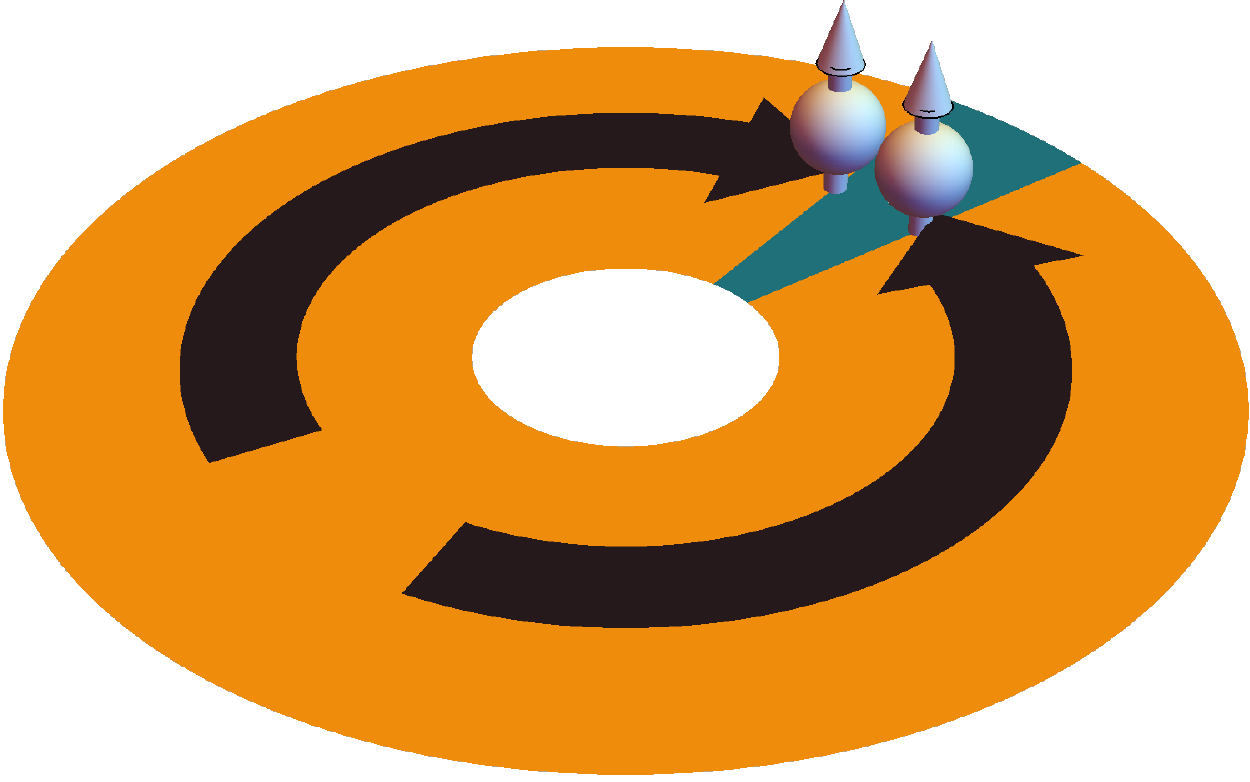}}
	\subfigure[]{\includegraphics[width=.4\columnwidth]{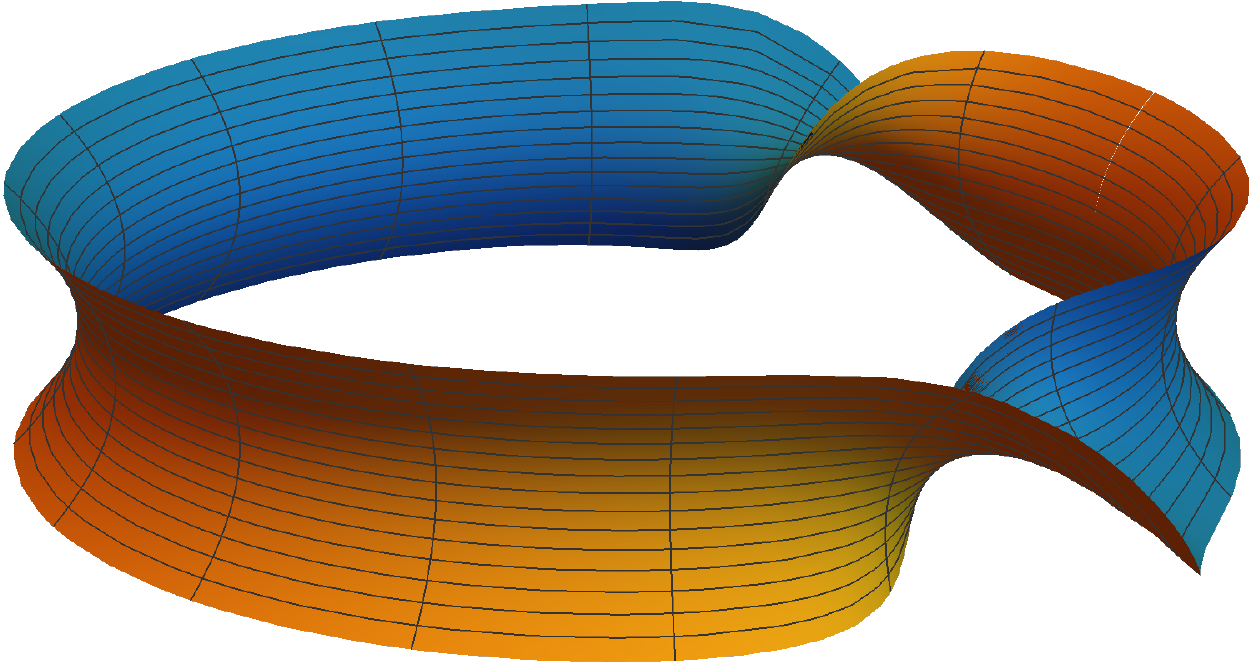}}
	\subfigure[]{\includegraphics[width=.4\columnwidth]{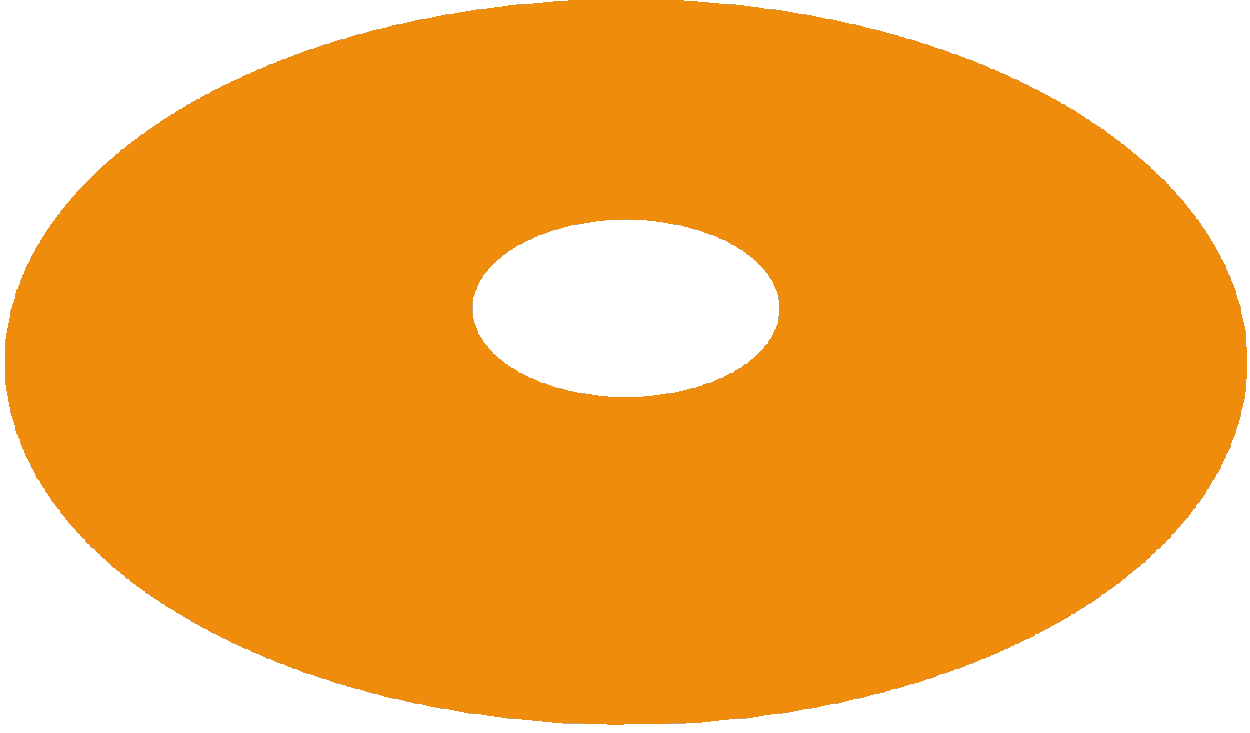}}
	\subfigure[]{\includegraphics[width=.4\columnwidth]{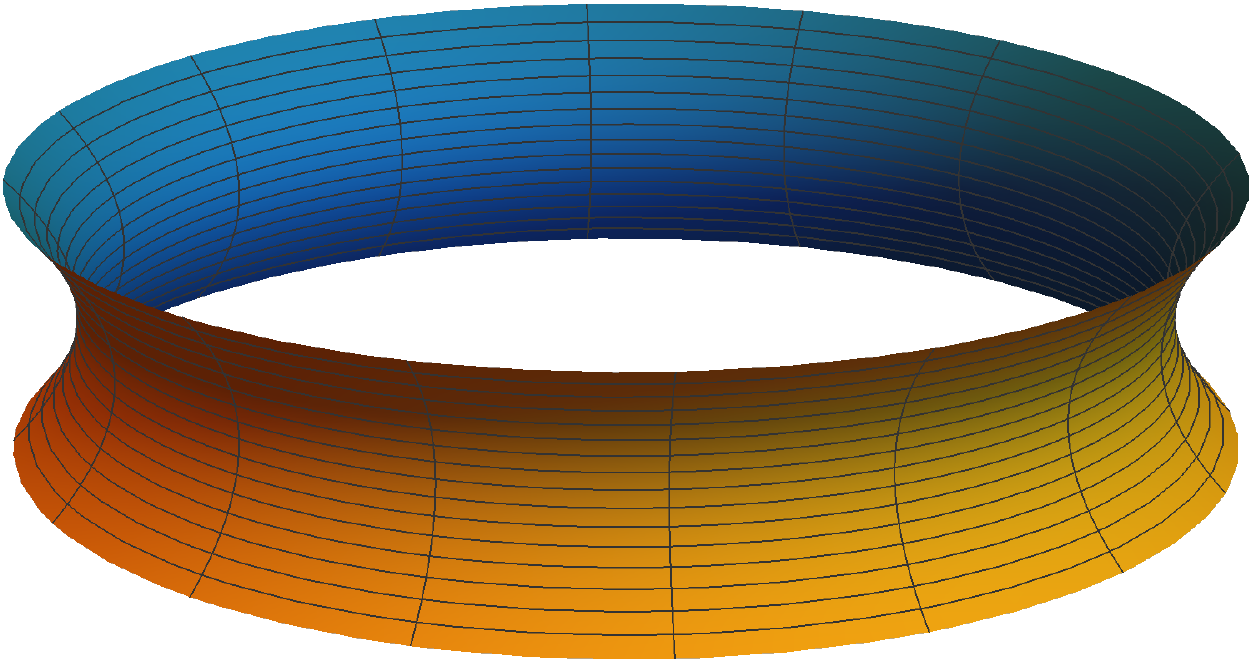}}
	\caption{Fractional excitations and topological degeneracy. The left column demonstrates a quantum $Z_2$ spin liquid (e.g. a RVB spin liquid) defined on an annulus, which has two degenerate ground states (a) and (g) due to topological degeneracy. Via local perturbation, a spin-1 excitation can be introduced to the first ground state, which can split into two spin-1/2 fractional excitations as show in Fig.~(c). If these two fractional excitations are moved around the annuls (e) and then annihilated with each other (g), the system turns into the other ground state, different from the original one where we start from. A catenoid with the same geometry setup shows the same property as show in the second column. Here, we also have two degenerate ground states, which correspond to swap the two sides of the 2D manifold. One can create two charge-1/2 solitons (d) and move them around the catenoid (f) before annihilate them. This procedure also flips a ground state into the other one.}
	\label{fig:spin:liquid}
\end{figure}

Here, instead of providing a comprehensive review about spin liquids in 2D, we examine a highly-simplified 1D version of dimer states, which exhibits a lot of the interesting ingredients of 2D spin liquids. Consider the one-dimensional Majumdar-Ghosh model~\cite{Majumdar1969},
which studies a 1D chain of spin-$1/2$ Heisenberg spins with frustrated nearest and next-nearest-neighbor anti-ferromagnetic couplings. This model has two degenerate ground states as shown in Fig.~\ref{fig:spin:chain}(a) and (b), where neighboring spins form singlet pairs and each spin can only participate in the formation of one spin singlet with one of its two neighbors. Such a singlet pair is called a ``dimer'' and the ground states of the 1D Majumdar-Ghosh model are called dimer states. This 1D chain is in strong analogy to
$Z_2$ spin liquids in 2D, whose ground states are also composed of dimerized  singlet pairs.
In the 1D Majumdar-Ghosh model, the two-fold ground-state degeneracy is due to spontaneous breaking of the lattice translational symmetry. 
For a $Z_2$ spin liquid defined on a 2D annulus  [Fig.~\ref{fig:spin:liquid} (a) and (g)], although the ground state breaks no symmetry, a similar 
two-fold degeneracy is expected due to topological reasons, which is known as topological degeneracy (see e.g. Ref.~\onlinecite{Wen2017} and references therein). In contrast to degeneracy from spontaneous symmetry breaking,
topological degeneracy implies that the number of degenerate ground states varies  according to the topology of 
the underlying manifold of the system in real space. A $Z_2$ spin liquid has $2^{n_L}$ ground states, where $n_L$ counts the number of 
independent non-contractible loops of the underlying manifold, e.g. a sphere or a disk has $n_L=0$. An annulus has $n_L=1$, while a torus or a double torus has $n_L=2$ or $n_L=3$ respectively. In this manuscript, to compare with narrow ribbons,
we focus on spin liquid defined on an annulus, with $n_L=1$ and thus $2$ degenerate ground states.

Now, we consider excitations in the Majumdar-Ghosh model. The obvious excitation in a dimerized ground state is to break a dimer, i.e. transfer a singlet into a triplet. Such a triplet excitation carries spin-1, which is a local excitation and can be introduced via a local perturbation. In the Majumdar-Ghosh model, such a local excitations can fractionalize into two spin-1/2 fractional excitations, as shown in Fig~\ref{fig:spin:chain} (c) and (d). Similar fractional excitations arise in 2D spin liquids. In both 1D and 2D, such a fractional excitation cannot be directly created in the bulk. Instead, they need to be created in pairs. More importantly, although each spin-1/2 excitation may look like an individual particle, fractional excitations in each pair are connected by a ``string'', which distinguish them form ordinary local excitations.

In a conventional material, the energy cost increases rapidly as one tries to split an integer-charged excitation into two parts and separate them away from each other.  
In a $Z_2$ spin liquid or a 1D Majumdar-Ghosh chain, however, the energy cost for separating two fractional excitations saturate quickly as their separation gets large and thus the attraction between the two fractional excitations decreases to zero as an exponential function of the separation. Therefore, once far apart, such a fractional excitation behaves just like a point particle, and is called a deconfined fractional excitation.
In addition to $Z_2$ spin liquids, this deconfinement mechanism is applicable to a wide variety of fractional excitations as well, such as fractional and non-abelian particles in fractional quantum Hall systems~\cite{Nayak2008}.

In topological states such as a Z$_2$ spin liquid, there exists one important connection between topological degeneracy and fractional excitations, known as braiding. 
This terminology usually refers to move a fractional particle around another one in 2D~\cite{Nayak2008}. However, similar phenomena often arise as long as 
the trajectory of a fractional particle forms a non-contractible loop, i.e. a closed loop which cannot smoothly shrink into a point, not necessarily due to the existence of another particle. Thus, in this paper, we will use this terminology loosely to refer to any non-contractible loops.
Imagine that we create a set of fractional excitations via certain local perturbation in a fractional topological state (e.g. a $Z_2$ spin 
liquid or a fractional quantum Hall system), and then adiabatically move their locations in a non-contractible way, i.e., the path of certain fractional excitation form a non-contractible closed loop. Afterwards, these fractional excitations are annihilated with one another and thus the system goes back to the ground state. Although this procedure starts from a ground states and ends also as a ground state, the initial and final states may be two distinct quantum states orthogonal to each other. 
One such example in a $Z_2$ spin liquid is demonstrated in Fig.~\ref{fig:spin:liquid}(a) (c) (e) and (g). 
As mentioned above, on a 2D annulus, this system has two degenerate ground states due to topological degeneracy. 
By creating a pair of fractional excitations and moving them around the annulus, the system is transformed from one ground state
to the other. In other words, braiding offers a pathway for topologically degenerate ground states to evolve into each other. This phenomenon plays a crucial role in topological quantum computing. If we consider two degenerate ground states as a quantum two-level system, these two states 
forms a q-bit and braiding serves as logical gates that flips and control this q-bit (For more details, see a review article Ref.~\onlinecite{Nayak2008} and references therein). 
For a Z$_2$ spin liquid, this braiding relation makes it possible to use this topological states as a topological quantum memory, where information stored in such a memory is robust against any local perturbations or quantum decoherence. The same principle can also be used for topological quantum computing. However, for such an objective, topological states with more complicated fractional particles and braiding algebra are needed, such as Majorana or Fibonacci anyons, where the later one can even achieve 
universal quantum computation~\cite{Freedman2002}.

This relation between degeneracy and fractional excitations also arises in 1D dimer states, and in Fig.~\ref{fig:spin:chain} (c) and (d), we can already see that moving
fractional particles flip the ground state (from blue to yellow).

Finally, it needs to be highlighted that their exists a deep connection between fractional excitations in a quantum spin chain and those in a minimal-surface narrow ribbon, 
if one realizes that low-energy physics in both systems are described the compact sine-Gordon theory. 
In the study of spin chains, one well-known and very powerful mathematical tool is the Luttinger liquid approach~\cite{Emery1979, 
Affleck1989, Francesco2012,Fradkin2013}. As shown by Haldane~\cite{Haldane1982}, in this approach, 
the low-energy effective theory of a dimerized spin chain is a compact sine-Gordon field theory, same as these ribbon systems. 
Remarkably, in the same work, Haldane also pointed out that spin-1/2 and spin-1 excitations there indeed correspond 
to charge-1/2 and charge-1 solitons respectively.
This analogy is the fundamental reason why these elastic ribbons share common properties with exotic quantum states formed by frustrated quantum spins.

\section{Fractional excitations in narrow ribbons}\label{SEC:FracRibb}
In this section, we discuss physical properties of fractional excitations in minimal-surface narrow ribbons.

\subsection{Integer and fractional solitons}
We first consider a minimal-surface narrow ribbon and assume that higher order terms in the elastic energy  favor the helicoid ground state.
As discussed above, key properties of such a ribbon can be characterized by the phenomenological theory of Eq.~\eqref{eq:sine_gordon} with $\gamma>0$. Most of our
conclusions can be easily generalized to other cases, e.g. for $\gamma<0$, where the ground state is a catenoid.

As shown in Fig.~\ref{fig:spin:chain} (e) and (f), this system has two degenerate ground states, i.e. left- or right- handed helicoids.
Now, we introduce soliton excitations to one of the ground states (e.g. right-handed) as shown in Fig.~\ref{fig:spin:chain} (g) and (h). 
With local deformations, only integer charged solitons can be introduced [Fig.~\ref{fig:spin:chain} (g)], which can fractionalize into two 1/2-charge fractional solitons
[Fig.~\ref{fig:spin:chain} (h)]. It is worthwhile to point out that for better visualization, we keep these helicoids straight even after solitons are introduced, whose 
true lowest-energy configuration should involve a kink as shown in Fig.~\ref{fig:numerical}.  
Same as the dimer chain [Fig.~\ref{fig:spin:chain} (a-d)], moving a pair of fractional solitons here flips one ground state into the other one.
Once created and becomes deconfined, a fractional soliton can no longer be removed by any local adjustment. Instead, it can only be neutralized by globally adjust 
the entire chain or by annihilating with an anti-soliton.

\begin{figure}[t]
	\subfigure[]{\includegraphics[width=.45\columnwidth]{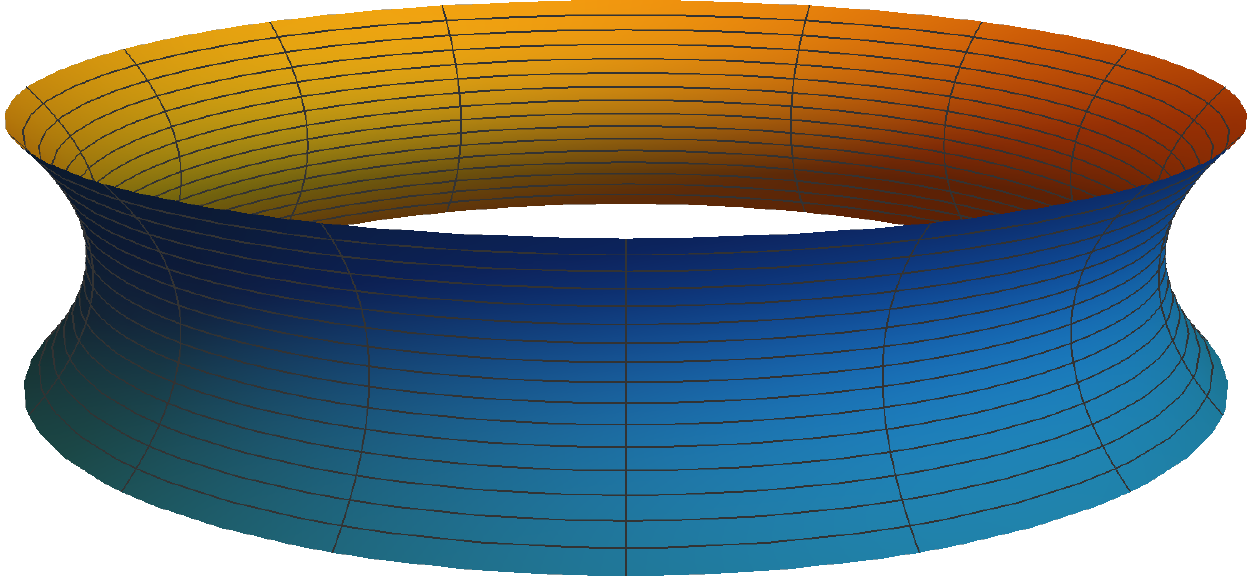}}
	\subfigure[]{\includegraphics[width=.45\columnwidth]{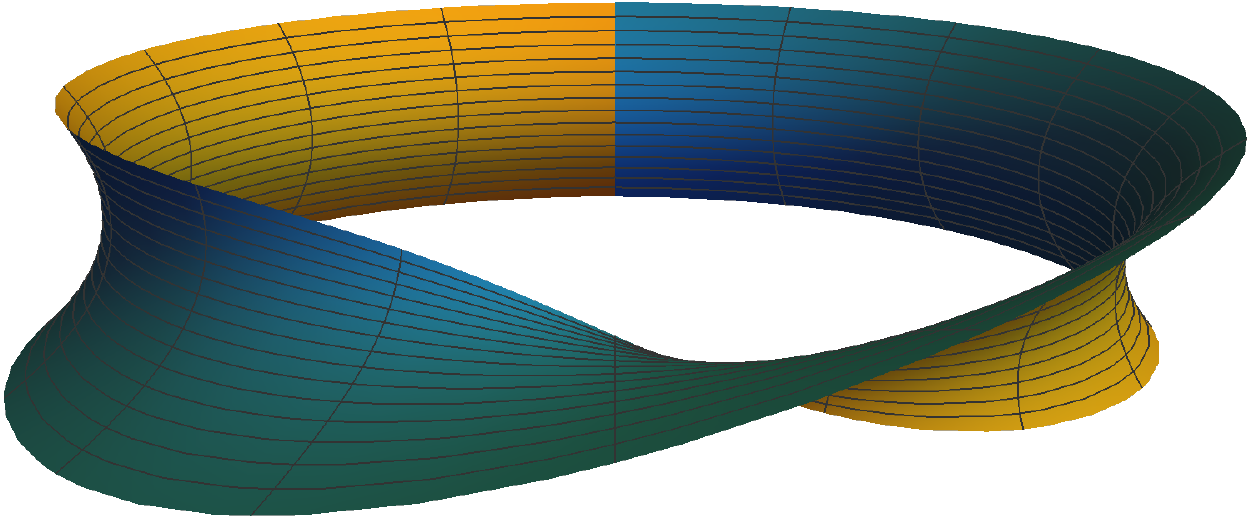}}
	\subfigure[]{\includegraphics[width=.45\columnwidth]{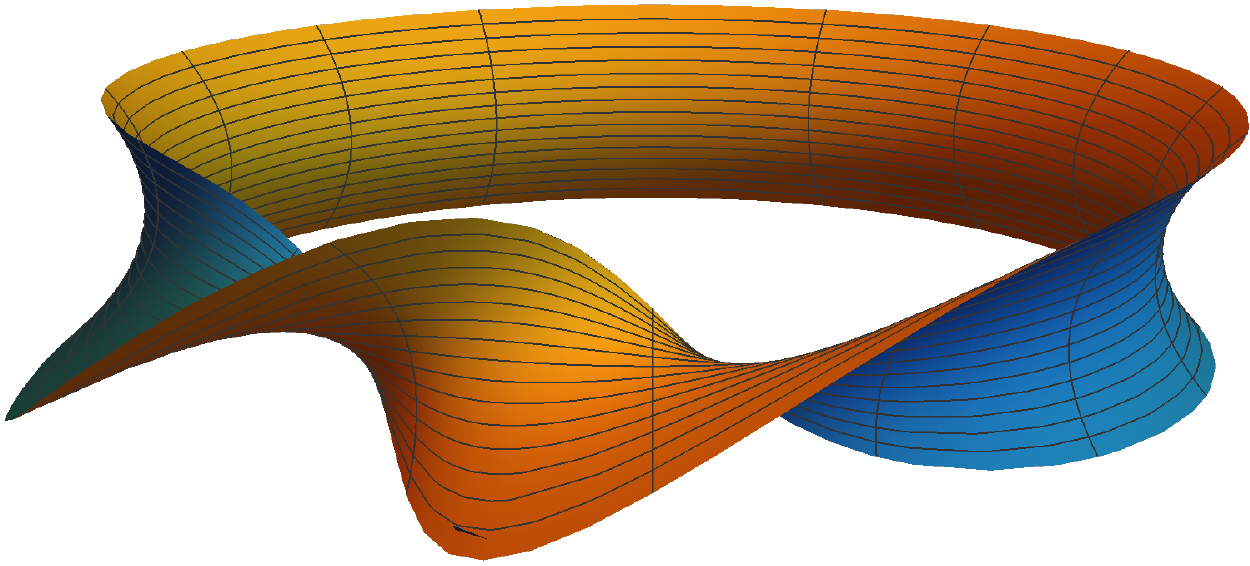}}
		\subfigure[]{\includegraphics[width=.45\columnwidth]{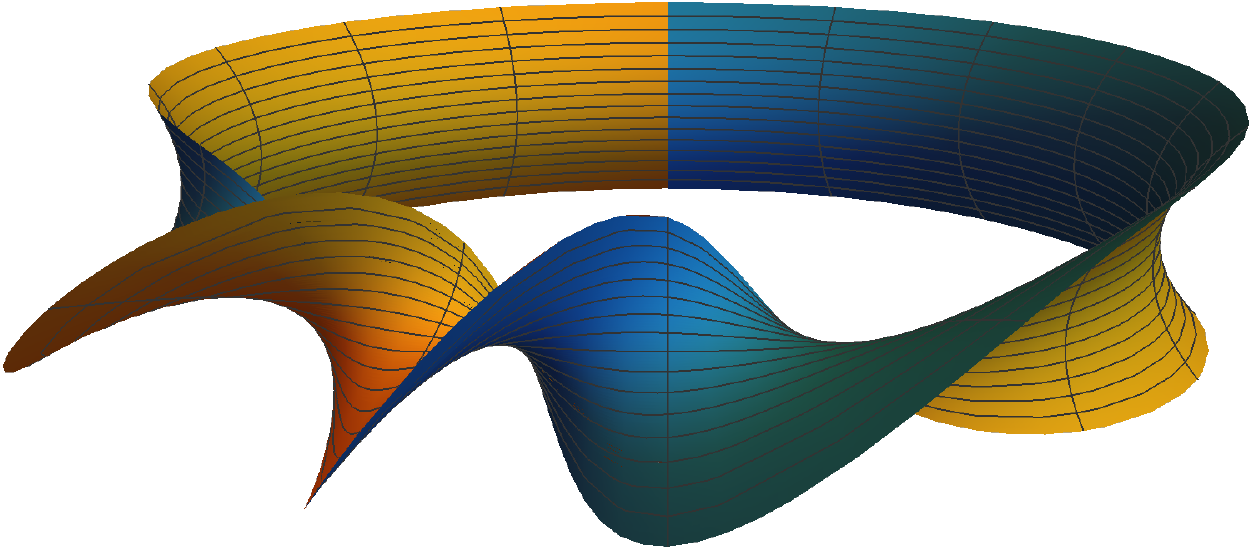}}
	\caption{Integer and fractional excitations in a catenoid. (a) shows a catenoid ground state without soliton excitations. (b-d) contain solitons. In contrast to an integer charged soliton (c), which preserves the orientability, a half-integer charged soliton makes the manifold non-orientable as shown in (b) and (d).}
	\label{fig:catenoid:solitons}
\end{figure}

In Fig.~\ref{fig:catenoid:solitons}, we consider a ribbon with catenoid ground states ($\gamma<0$) and assume that the catenoid form a closed loop as shown in the figure. 
Interestingly, creating a half-integer-charge fractional soliton must sacrifice the orientability, turning the orientable cylinder-like structure into an non-orientable surface.
which offers another hint about the fractional nature of these solitons.

\subsection{Two unique features}
\label{sec:sub:two:features}
Fractional solitons in a narrow helical ribbon have two unique features, which are not generally expected for most other fractional excitations.
Firs of all, these solitons are \emph{holographic}, which means that if there is only one charge-1/2 soliton in a helicoid, we can pin-point and control its location via 
controlling the two ends of the helicoid.
This is because the soliton here is the domain boundary between left- and right- handed sections. For a helicoid with length $L$, we define the left-handed section length to be $x$, and thus the right-handed section shall have length $L-x$. For simplicity, here we ignore the width of the soliton, which can be included easily without alternating any main conclusions. For a helicoid, we can define the total helicity of the entire ribbon as $(L-2x)/\lambda$ where $\lambda$ is the pitch of the helicoid. This quantity describes how many times the ribbon twists,
with right-handed twists defined as positive. It is easy to notice that this quantity directly connects helicity with the location of the fraction soliton $x$. For an elastic ribbon, the helicity can be adjust by twisting the two ends of the ribbon in the opposite direction and each $2\pi$ twist increases/decreases the helicity by $1$, which moves the fractional soliton by one 
pitch $\lambda$. This holographic control is not a general property of fractional solitons, but a special feature for solitons in helicoids. 
In addition to control fractional solitons, the holographic property also provide a natural way to generate these fractional excitations. If we twist the two ends of a helicoid such that 
the helicity decreases from the ground state value ($L/\lambda$), this process will eventually create a fractional soliton (i.e. a non-zero $x$) to reduce energy. As will be discussed 
in the next section, this is the mechanism how solitons in telephone cords is generated.

The second feature of these fractional solitons is that they kink the helical ribbon by turning its direction by almost $180^\circ$ as discussed in Sec.~\ref{SEC:Nume}. Again, this is a special feature, not generally expected for fractional excitations.
For a telephone cord, this is the reason why solitons  tangle up the cord.

\subsection{Braiding}
In this section we demonstrate how braiding changes the ground state in narrow ribbons by considering a catenoid ground state $g<0$. As mentioned early on, here
we are using this terminology loosely, which includes any non-contractable loop trajectory.

Here, we consider one orientable cylinder-like catenoid [Fig.~\ref{fig:spin:liquid}(b)] and introduce a pair of fractional solitons by locally flipping a section of 
the the catenoid inside out [Fig.~\ref{fig:spin:liquid}(d)]. Then, we gradually move the pair of fractional solitons around the catenoid [Fig.~\ref{fig:spin:liquid}(f)], where they meet
again and annihilate with each other. It is easy to realize that such a procedure changes the ground state, in analogy to the same procedure in a $Z_2$ spin liquid. 

It is worthwhile to mention that here, as we annihilate two solitons, large dissipation is assumed, such that a system can quickly release its elastic energy and dissipate from an excited state (with a pair of solitons) to  the ground state (without solitons). In the absence of dissipation, the system will remain in excited states and thus solitons cannot be annihilated, even if a pair of solitons  collide. In such a scenario, the solitons will pass through each other, which is in fact part of the definition of solitons.

\section{discussion}
Now we come back to telephone cords. 
A helical telephone cord is not a thin 2D helicoid ribbon, but we can consider it as a stack of 2D helicoid ribbons and thus they are expected to share similar qualitative features
as our minimal-surface ribbons.  The insight we gain from the discussions above on fractional solitions provide us with answers to the three questions we raised in the Introduction.

First, when we use a telephone headset, we often unintentionally twist/rotate the headset before we put it back. As shown in Sec.~\ref{sec:sub:two:features}, 
due to the holographic property, this type of twisting introduces fractional solitons.  This is the origin, from which these solitons emerge in  phone cords.
Second, in Sec.~\ref{sec:sub:two:features}, we showed that such a soliton bend the ribbon, which is why they result in kinks in phone cords and turn them into a tangled mess. 
Third, because these fractional excitations are topological defects, they cannot be removed by any local deformations, and this is the reason why these annoying kinks are 
so hard to get rid off.
With the origin of the kinks understood, we can use this knowledge as guidance to remove/avoid these kinks. By avoiding rotating the headset, these kinks would not be 
created. For telephone cords that already have these kinks, we just need to twist the headset, which moves the solitons, making them annihilate with each other or driving them out
of the cord.

Similar types of kinks/solitons between domains of helical structures with opposite handedness have been observed previously in various situations such as perversion of tendrils on climbing plants~\cite{Goriely1998}, intrinsically curved rods~\cite{domokos2005multiple}, self-assembled structures of Janus colloidal particles~\cite{Chen2011supra},
elastic bi-strips~\cite{liu2014structural}, helical strings~\cite{Nisoli2015}, and minimal surface liquid films~\cite{Machon2016}.  In this paper we show how this type of solitons arises  in narrow elastic ribbons and are fractional excitations flipping the ribbon between ground states that belong to the same minimal surface associate family.  We further demonstrate that they share the same topological description as fractional excitations in $Z_2$ spin liquids.

The fractional soliton excitations we discussed in the helicoid-catenoid family are protected by the chiral symmetry of the elastic theory.  Interestingly, when this symmetry is weakly broken, e.g., the left- and right- handed helicoids have slightly different elastic energy but both remain local energy minima, the soliton propagation between these two domains will have a preferred direction, such that it flips the higher energy domain into the lower energy domain.  Interestingly, because this soliton is holographic, its propagation requires the rotation of the domains, the sizes of which change as the soliton propagates.  
This results in a locked speed for the soliton propagation, as the released elastic energy from the domain-energy difference 
  turns into kinetic energy and fuels the rotation of the domain (the size of which increases).  This mechanism of locked soliton speed was first discussed in the context of solitons in helical strings, in Ref.~\onlinecite{Nisoli2015}.  In contrast, if the chiral symmetry is preserved and the two domains have exactly the same elastic energy, the speed of the soliton changes at it propagates, because the domain size changes but the total energy is conserved.  This phenomenon of locked soliton speed may have broad applications as a constant actuation/propulsion mechanism without the need of fine-tuning.

Topological solitons have also been found in discrete 1D rotor chains, which are  Maxwell lattices with exactly one floppy mode under open boundary conditions~\cite{Chen2014}.  Interestingly, solitons in these 1D chains exhibit a kink-antikink asymmetry~\cite{Zhou2017}.  
If we follow the same soliton charge defined above, this asymmetry implies the breaking of the charge conjugation symmetry for such solitons, i.e., positively and negatively charged solitons are no longer equivalent,  in sharp contrast to conventional solitons. 
Whether similar type of excitations can arise in continuum would be an interesting question for future studies.

In addition to narrow ribbons, similar physics of associate family arises in any minimal-surface 2D plates and their low-energy properties share the same sine-Gordon
description, from which fractional excitations can also arise. Generalizing this knowledge about fractional excitations to other 2D plates will be an interesting subject for future studies,
from which a universal understanding about fractional excitations in systems at the verge of mechanical stability may eventually emerge. 

\appendix

\section{2D manifolds embedded in a 3D Euclidean space}
\label{app:sec:2Din3D}
In this section, we summarize the mathematical tools that are needed to characterize 2D manifolds embedded in a 3D space. 

\subsection{The Einstein notation}
For curved Riemannian manifolds, covariance and contravariance vectors (tensors) need to be distinguished. In this manuscript, we follow the Einstein notation 
to represent them by lower- and upper- indices respectively. 
These covariance and contravariance variables are related with each other via raising/lowering the index through the metric tensor $g$. 
We define $g^{ij}$ as the inverse of $g_{ij}$ and the trace of a rank two tensor is  $\tr A= A_{ij}g^{ij}$ and $\tr A^2= A_{ij}g^{jk} A_{kl}g^{li}$.  

\subsection{The first-fundamental form and the Gaussian curvature}
For a smooth 2D manifold, its metric tensor (i.e. the first fundamental form) takes the following form. 
\begin{align}
g=
\begin{pmatrix}
E(u,v) & F(u,v)\\
F(u,v) & G(u,v)
\end{pmatrix}
\end{align}
where $(u,v)$ is the 2D coordinate and the matrix element $E$, $F$ and $G$ are functions of $u$ and $v$. 

The first fundamental form not only defines the measurement of ``distance'', but also uniquely determines the Gaussian curvature as
\begin{align}
K=-\frac{1}{E} (\partial_u \Gamma^{2}_{12} - &\partial_v \Gamma^{2}_{11}
+\Gamma^1_{12}\Gamma^2_{11}\nonumber\\
&
-\Gamma^1_{11}\Gamma^2_{12}
+\Gamma^2_{12}\Gamma^2_{12}
-\Gamma^1_{11}\Gamma^2_{22})
\label{app:gauss_curvature_1}
\end{align}
where $\Gamma$ is the Christoffel connection
\begin{align}
\Gamma^i_{jk}=\frac{g^{i l}}{2}(\partial_k g_{lj}+\partial_j g_{lk}-\partial_l g_{jk})
\end{align}
with $i$, $j$ and $k$ being $1$ or $2$.

Here, it is worthwhile to emphasize that as shown in Eq.~\eqref{app:gauss_curvature_1}, the Gaussian curvature is
uniquely pined by the first fundamental form. As will be shown below, this enforces one important constraint on the second 
fundamental form (and fixes its determinant).

\subsection{The second-fundamental form and curvatures}
The second fundamental form $h$ is also a $2\times 2$ symmetric matrix, whose components are
\begin{align}
h_{i,j}=\mathbf{n}\cdot \partial_i \partial_j \mathbf{R}
\end{align}
where $\mathbf{n}=(\partial_u \mathbf{R}\times  \partial_v \mathbf{R})/|\partial_u \mathbf{R}\times  \partial_v \mathbf{R}|$ is the unit vector along the normal direction.
Here $\mathbf{R}$ is the 3D coordinate of the target space.

In general, $h$ takes the following structure
\begin{align}
h=
\begin{pmatrix}
L(u,v) & M(u,v)\\
M(u,v) & N(u,v)
\end{pmatrix}
\end{align} 
with $L=\partial_u^2\mathbf{R}\cdot \mathbf{n}$, $N=\partial_v^2\mathbf{R}\cdot \mathbf{n}$ and $M=\partial_u\partial_v\mathbf{R}\cdot \mathbf{n}$

The second fundamental form describes the curvatures of a 2D manifold, where the Gaussian curvature $K$ and the mean curvature $H$ are
\begin{align}
K&=\frac{\det h}{\det g}
\label{app:gauss_curvature_2}
\\
H&=\frac{1}{2 \det g}(G L+E N-2 F M)
\end{align}
with $\det$ representing the determinant.
It is worthwhile to emphasize that Eq.~\eqref{app:gauss_curvature_2} implies  
\begin{align}
K\times \det g = \det h .
\label{app:gauss_curvature_constraint}
\end{align}
The l.h.s. of this equation
is fully dictated by the first fundamental form [via Eq.~\eqref{app:gauss_curvature_1}], while the r.h.s. only relies on the second  fundamental form.
Thus, this equation implies that the first fundamental form, once determined, shall fully dictates the determinant of the second fundamental form.

\subsection{Gauss-Codazzi equations}
The first- and second- fundamental forms of a 2D manifold must satisfy two (sets of) constraints.
The first constraint is from the two definitions of the Gaussian curvature [Eqs.~\eqref{app:gauss_curvature_1} and~\eqref{app:gauss_curvature_2}] , which must produce the same value. As shown in Eq.~\eqref{app:gauss_curvature_constraint}, this constraint enforce a nonlinear relation between the first and second fundamental form.
In addition,  the first- and second- fundamental form must satisfy a set of partial differential equations, known as the Gauss-Codazzi equations, which are
\begin{align}
\partial_v L -\partial_u M &=L \Gamma^1_{12}+M (\Gamma^2_{12}-\Gamma^1_{11})-N \Gamma^2_{11}\\
\partial_v M -\partial_u N &=L \Gamma^1_{22}+M (\Gamma^2_{22}-\Gamma^1_{21})-N \Gamma^2_{21}
\end{align}

\subsection{Orthogonal coordinates}
In a neighborhood of any non-singular point of a smooth 2D manifold, an orthogonal coordinate always exists, under which the metric tensor is diagonal ($F=0$). In this manuscript, without loss of generality, we will use orthogonal coordinates and thus $F$ can always be set to zero, 
\begin{align}
g=
\begin{pmatrix}
E(u,v) & 0\\
0 & G(u,v)
\end{pmatrix}
\end{align}
In this coordinate, Eq.~\eqref{app:gauss_curvature_1} takes a simpler form
\begin{align}
K=\frac{1}{4E G}\left(\frac{E_v^2+E_u G_u}{E}+\frac{G_u^2+E_vG_v}{G}-2 E_{vv}-2 G_{uu}\right)
\label{app:gauss_curvature_ortho}
\end{align}
where subindex $u$ and $v$ represent partial derivatives, e.g. $E_v=\partial_v E$ and $E_{vv}=\partial_v^2 E$ and the Christoffel connection takes the form of
\begin{align}
&\Gamma^{1}_{11}=\frac{E_u}{2E} \\
&\Gamma^{2}_{11}=-\frac{E_v}{2G} \\
&\Gamma^{1}_{12}=\frac{E_v}{2E} \\
&\Gamma^{2}_{12}=\frac{G_u}{2G} \\
&\Gamma^{1}_{22}=-\frac{G_u}{2E} \\
&\Gamma^{2}_{22}=\frac{G_v}{2G}
\end{align}

In an orthogonal coordinate ($F=0$), these formula can be simplified by introducing the matrix
\begin{align}
\tilde{h}=
\begin{pmatrix}
l(u,v) & m(u,v)\\
m(u,v) & n(u,v)
\end{pmatrix}
\label{app:h_tilde}
\end{align}
where $l=L/E$, $n=N/G$ and $m=M/\sqrt{E G}$. With the $\tilde{h}$ matrix, the Gaussian curvature becomes
 \begin{align}
K&=\det \tilde{h} = l n-m^2
\label{app:gauss_curvature_3}
\\
H&=\tr \tilde h/2= (l+n)/2
\label{app:mean_curvature_2}
\end{align}

If we choose an orthogonal coordinate, the Gauss-Codazzi equations shall take the form
\begin{align}
\partial_v L -\partial_u M &=L \frac{E_v}{2E} +M (\frac{G_u}{2G}-\frac{E_u}{2E})+N \frac{E_v}{2G}\\
\partial_v M -\partial_u N &=- L \frac{G_u}{2E}+M (\frac{G_v}{2G}-\frac{E_v}{2E})-N \frac{G_u}{2G}
\end{align}
In terms of $l$, $m$ and $n$, these equations are
\begin{align}
\partial_v l - \sqrt{\frac{G}{E}} \partial_u m &=(n-l) \frac{E_v}{2E} +m \frac{G_u}{\sqrt{G E}}
\label{app:Codazzi_l_m_n_1}\\
\partial_u n -\sqrt{\frac{E}{G}} \partial_v m &=(l-n) \frac{G_u}{2G} + m \frac{E_v}{\sqrt{G E}}
\label{app:Codazzi_l_m_n_2}
\end{align}

\section{Stress-free 2D plates}
\label{app:sec:stress:free:2D}
In this section, we utilize the mathematical tools discussed in the previous section to find conditions, under which a 2D plate with $E=E_s+E_b$ is stress free.

We follow the same strategy outlined in the main text. First, we minimize the dominant part of the elastic energy $E_s$, which requires $g=g_0$ and thus pin down the first fundamental form ($g$) and the Gaussian curvature ($K$). Then we minimize $E_b$ with fixed $g$ and $K$. In order for the plate to stay stress free, this minimization also need to satisfy the Gauss-Codazzi equations.

\subsection{$K>0$}
From Eq.~\eqref{app:gauss_curvature_3}, it is easy to realize that $l n>0$ if $K>0$. Utilizing Eq.~\eqref{app:mean_curvature_2}, we get
\begin{align}
H^2= (l+n)^2/4 \ge  l n=K+m^2 \ge K
\end{align}
Here, we utilized the inequality of arithmetic and geometric means, as well as the fact that $m^2 \ge 0$. Thus for $K>0$, $H^2$ cannot be smaller than $K$ and 
this minimum value of $H^2$ is reached only at $m=0$ and $l=n=\sqrt{K}$.

Now, we check whether and when this energy minimum can satisfy the Gauss-Codazzi equations. With $m=0$ and $l=n=\sqrt{K}$, Eqs.~\eqref{app:Codazzi_l_m_n_1}
and~\eqref{app:Codazzi_l_m_n_2} take the following form 
\begin{align}
\partial_v l =0\;\; \textrm{and} \;\; \partial_u n=0
\end{align}
Because $l=n$, these two equations imply that $\partial_u l=\partial_v l=0$ and the same is true for $n$, i.e., $l=n=\sqrt{K}$ is a constant independent of $u$ and $v$. 
Obviously, this means that for $K>0$, this energy minimum satisfies the Gauss-Codazzi equations, if and only if the manifold has a constant Gaussian curvature.
For a manifold with constant positive Gaussian curvature, $l=n=\sqrt{K}$ and $m=0$ imply that this stress-free ground state must be (part of) a sphere.

\subsection{$K<0$}
For $K<0$, the minimum of $H^2$ is zero and this minimum is reach whenever $l=-n$. In terms of the $\tilde{h}$ matrix defined in Eq.~\eqref{app:h_tilde}, this means that $\tilde{h}$ 
is traceless (and by definition the determinant of this matrix is $K$). Thus, at this energy minimum the $\tilde{h}$ matrix must take the following structure
\begin{align}
\tilde{h}=& \sqrt{-K} 
\begin{pmatrix}
-\cos\varphi & \sin \varphi\\
\sin \varphi & \cos\varphi
\end{pmatrix}\nonumber\\
= & \sqrt{-K}  (\sin \varphi \; \sigma_x-\cos \varphi \; \sigma_z)
\end{align}
where $\sigma_x$ and $\sigma_z$ are two of the Pauli matrices. The angle $\varphi$ is
the same phase angle $\varphi$ that we used in the main text to define the mechanism. It is easy to verify that this $\tilde{h}$ is traceless (and thus $H=0$) and it produces the correct Gaussian curvature $K$.
In contrast to the $K>0$ case, where one unique minimum is found, here we have a free parameter $\varphi$ and thus infinitly many degenerate ground states.

Now, we need to check whether and when these minimum-energy  configurations are allowed by the Gauss-Codazzi equations. In other words, for any arbitrary choice of $g_0$, whether 
the first fundamental form $g=g_0$ and  the second fundamental form which has $H=0$ obey the Gauss-Codazzi equations. As will be shown below, the $H=0$ condition is
equivalent to requiring that the 2D manifold is a minimal surface. Thus, the question of whether the Gauss-Codazzi equations are obeyed here is fully equivalent to the following question: for a given $g_0$, whether there exists an isometric embedding such that $g=g_0$ and the manifold is a minimal surface. The answer to this question is known and has been proved~\cite{Ricci1894}, which is the minimal surface condition shown in the main text in Sec.~\ref{SEC:PlatBend}.

\section{Minimal surfaces}
\label{app:sec:minimal:surface}
In this section, we provide a brief review about basic concepts and properties of minimal surfaces. Minimal surfaces are 2D surfaces that locally minimize their area, which is equivalent to requiring these surfaces to have zero mean curvature.

\subsection{Weierstrass-Enneper parameterization}
Mathematically, minimal surfaces have a deep and fundamental connection with complex analysis. 
It is know that all minimal surfaces can be represented using the Weierstrass-Enneper parameterization
\begin{align}
R_1&=\Re  \int \mathfrak{f}  (1-\mathfrak{g}^2)/2 \;\; dz \\
R_2&= \Re \int i \mathfrak{f} (1+\mathfrak{g}^2)/2 \;\; dz  \\
R_3&=\Re \int \mathfrak{f \; g} \;\;dz
\end{align}
where $\Re$ represents the real part. $\mathbf{R}=(R_1, R_2, R_3)$ is the 3D coordinate of the target space, while the 2D coordinate of the original space (i.e. the 2D manifold) $\mathbf{r}=(x,y)$ is represented by the  complex variable $z=x+i y$. $\mathfrak{f}$ and $\mathfrak{g}$ are complex functions of $z$, where $f$ is holomorphic and $g$ is meromorphic. In complex analysis, holomorphic means that a function is analytic with well defined Taylor expansions for every point in a domain, while meromorphic is a slightly weaker condition, which is similar to an analytic function but can contain a set of isolated singular points (poles). It is easy to verify that the 2D manifold defined by $\mathbf{R}(\mathbf{r})$ has zero mean curvature and thus is a minimal surface.

In the Weierstrass-Enneper parameterization, an associate family is represented by a phase factor $e^{i\varphi}$. As can be easily verified, by multiplying a constant phase factor to the function $\mathfrak{f}$ , we obtain a family of minimal surfaces via the Weierstrass-Enneper parameterization  
\begin{align}
R_1&=\Re\; e^{i \varphi} \int \mathfrak{f}  (1-\mathfrak{g}^2)/2 \;\;dz \\
R_2&= \Re\; e^{i \varphi}\int i \mathfrak{f} (1+\mathfrak{g}^2)/2 \;\;dz  \\
R_3&=\Re \;e^{i \varphi} \int \mathfrak{f\; g} \;\;dz
\end{align}
All minimal surfaces in this associate family share the same metric tensor $g$ and the same mean curvature ($H=0$), and they can be evolved smoothly into each other via 
adiabatically varying the value of $\varphi$. Because our elastic energy $E=E_s+E_b$ only depends on $g$ and $H$, for a 2D plate with minimal-surface 
ground state, all configurations in the associate family are degenerate ground states and there exits a floppy mode to deform these ground states into each other smoothly without energy cost.

From equations shown above, it is easy to realize that under the transformation $\varphi \to \varphi+\pi$, $\mathbf{R}\to-\mathbf{R}$, which flips the chirality. 

\subsection{The helicoid-catenoid family}
For minimal surfaces in the helicoid-catenoid family, the Weierstrass-Enneper parameterization takes the following form
\begin{align}
\mathfrak{f}= i e^{-z} \;\;\; \mathrm{and}\;\;\; \mathfrak{g}=-i e^{z}
\end{align}
Thus
\begin{align}
R_1&=- \cosh x \sin y \cos \varphi - \sinh x \cos y \sin\varphi \\
R_2&= \cosh x \cos y \cos \varphi- \sinh x \sin y \sin\varphi   \\
R_3&=x \cos\varphi - y \sin \varphi 
\end{align}
Under a coordinate transformation $u=\sinh x$ and $v=y$, we get
\begin{align}
R_1&=-\sqrt{1+u^2} \sin v \cos \varphi - u \cos v \sin\varphi \\
R_2&=\sqrt{1+u^2} \cos v \cos \varphi - u \sin v  \sin\varphi \\
R_3&= \arcsinh u \cos\varphi - v \sin \varphi 
\end{align}
For $\varphi=\mp \pi/2$, we have helicoids with left/right handness
\begin{align}
\mathbf{R}=\pm(u \cos v, u \sin v,v)
\end{align}
For $\varphi=0$ or $\pi$, catenoids are obtained with
\begin{align}
\mathbf{R}=\pm (-\sqrt{1+u^2} \sin v, \sqrt{1+u^2} \cos v, \arcsinh u)
\end{align}
Other values of $\varphi$ gives other minimal surfaces in this associate family (Fig.~\ref{fig:associate:family}).

\section{Narrow ribbons}
\label{app:sec:ribbons}
For a ribbon, we choose the direction of $v$ to be along the ribbon, while $u$ is along the perpendicular direction and $-w/2\le u\le w/2$ with $w$ being the width of the ribbon. 
Because a ribbon preserves the transitional symmetry along $v$, i.e. a ribbon is invariant under $v\to v+ \delta v$ for any arbitrary real $ \delta v$, the metric tensor $g_0$ must be independent of $v$, and thus is a function of $u$ only. In addition, By choosing an orthogonal coordinate and a simple rescaling, the metric tensor can always be written as
\begin{align}
g_0=
\begin{pmatrix}
1 & 0\\
0 & g_{22}(u)
\end{pmatrix}
\label{app:g0_general}
\end{align}
For a narrow ribbon (small $w$), one can expand $g_{22}(u)$ as a power law series
\begin{align}
g_{22}(u)= b_0 + b_1 u +b_2 u^2 +O(u^3)
\label{app:g22_expansion_0}
\end{align}
Here, we ignore cubic and higher order terms. This approximation is valid in the narrow limit, where $|u|\le w/2$ is small. Higher order corrections here are of the order $O(w^3)$ in $g_0$. For the elastic energy $E_s$, the higher terms ignored here leads to $O(w^4)$ corrections, with one extra power coming from the spatial integral. 
The constant term in $g_{22}$, i.e., the first term in Eq.~\eqref{app:g22_expansion_0}, can always be set to identity by a rescaling $v \to \sqrt{b_0} v$, and thus we have 
\begin{align}
g_{22}(u)= 1 + a_1 u +a_2 u^2 +O(u^3)
\label{app:g22_expansion}
\end{align}
with $a_1=b_1/b_0$ and $a_2=b_2/b_0$.

If $ a_1^2/4<a_2$, it is easy to check that the Gaussian curvature of this $g_0$ is negative
\begin{align}
K_0=-\frac{a_2-a_1^2/4}{1+a_1 u+a_2 u^2}<0
\label{app:curvature_ribbon}
\end{align}
Following the minimal-surface condition defined in the main text, we now treat $\sqrt{-K_0} g_0$ as a new metric tensor
\begin{align}
g_1=\sqrt{-K_0} g_0=
\begin{pmatrix}
\frac{\sqrt{a_2-a_1^2/4}}{1+a_1 u+a_2 u^2} & 0\\
0 & \sqrt{a_2-a_1^2/4}
\end{pmatrix}
\end{align}
and compute  the Gaussian curvature that this metric tensor $g_1$ defines $K_1$.  Using Eq.~\eqref{app:gauss_curvature_ortho},
it is easy to check that $K_1=0$, and thus the minimal surface condition is satisfied as long as $a_2> a_1^2/4$, regardless of microscopic details.
As will be shown in the next section, the ground state here is in the helicoid-catenoid minimal-surface associate family. 

For $a_1^2/4 \ge a_2$, the Gaussian curvature of $g_0$ is non-negative [Eq.~\eqref{app:curvature_ribbon}]. Up to small corrections of the order $O(u^3)$ or higher, 
the metric tensor $g_0$ can be rewritten as
\begin{align}
g_0=
\begin{pmatrix}
1 & 0\\
0 & \frac{\cos^2 (u/\rho_0-\phi_0)}{\cos^2 \phi_0}
\end{pmatrix}+O(u^3)
\label{app:approximate_g0_sphere}
\end{align}
where the two constants are $\rho_0=1/\sqrt{a_1^2/4-a_2}$ and $\phi_0=\arctan(a_1 \rho /2)$. 
It is easy to check that this $g_0$ [Eq.~\eqref{app:approximate_g0_sphere}]  is identical to the original $g_0$ [defined in Eqs.~\eqref{app:g0_general} and~\eqref{app:g22_expansion}] up to small corrections of  the order $O(u^3)$,  which are ignored in the thin ribbon limit as shown above.
Utilizing Eq.~\eqref{app:gauss_curvature_ortho}, it is easy to check that this metric tensor [Eq.~\eqref{app:approximate_g0_sphere}] results in a constant positive Gaussian curvature $K=1/\rho^2=(a_1^2/4-a_2)$. In other words, for any thin ribbons with $a_1^2/4 \ge a_2$, the sphere condition is satisfied and thus we have a stress-free ground state, which is part of a sphere with radius $\rho_0=1/\sqrt{a_1^2/4-a_2}$.

In summary, regardless of microscopic details, a narrow ribbon can always satisfy one of the two stress-free conditions and thus reach a stress-free ground state, which is either part of a sphere or part of a minimal surface in the helicoid-catenoid associate family.

\section{Higher order terms}
\label{app:sec:higher:order}
As mentioned in the main text, for a 2D plate that satisfies the minimal-surface criterion, the system has a floppy mode and thus infinitelt many degenerate ground states, 
i.e. all minimal surfaces in the corresponding associate family. In a real material, such a infinite ground-state degeneracy will in general be lifted by 
higher order terms in the elastic energy. 
For a ribbon, this will result in two degenerate ground states, connected with each other by the $\varphi \to \varphi+\pi$ transformation as shown
in the main text. In this section, we demonstrate one example of such higher order terms, which favor the helicoid- or catenoid- ground states.

As shown in Eq.~\eqref{eq:Eb}, for an isotropic (or nearly isotropic) material, the bending energy depends on the the mean curvature $H$ and the Gaussian curvature $K$.
Using the $\tilde{h}$ matrix defined in Eq.~\eqref{app:h_tilde}, the bending energy [Eq.~\eqref{eq:Eb}] can be written as
\begin{align}
E_b=&\int dr{\sqrt{\det g_0}}\left[ D_1 H^2 - D_2  K \right] \nonumber\\
=&\int dr{\sqrt{\det g_0}}\left[ \frac{D_1}{4} (l+n)^2 - D_2  (l n-m^2) \right]
\label{eq:Eb:app}
\end{align}
where $D_1$ and $D_2$ are coefficients, whose values are shown in Eq.~\eqref{eq:Eb}. 
Here, we used the fact that $H=\tr{\tilde{h}}/2$ and $K=\det \tilde{h}$ shown in Eqs.~\eqref{app:gauss_curvature_3} and \eqref{app:mean_curvature_2}.
This bending energy is isotropic, i.e., the energy cost is identical no matter we bend along the main axis ($u$ or $v$) or the diagonal direction ($u+v$ or $u-v$).

For a real 2D plate with $D_{4h}$ symmetry (e.g. materials with a tetragonal lattice) or lower symmetries, 
the bending term is no longer isotropic, and thus extra terms become allowed, such as
\begin{align}
\delta E_b=\int dr{\sqrt{\det g_0}} \left[   \frac{\delta D}{4} (l^2+n^2)\right]
\label{eq:deltaEb:app}
\end{align}
where $\delta D$ is a coefficient, which describes the anisotropy between the main-axis and the diagonal directions. 
In this manuscript, we focus on nearly isotropic systems, and thus we will always assume that $\delta D \ll D_1$ and   $\delta D \ll D_2$.
For positive (negative) $\delta D$, this term implies that it is harder (easier) to bend the plate along the main-axis direction, in comparison to diagonal.
With $\delta D>0$, this term lifts the infinite degeneracy of the narrow ribbon and selects two configurations as the real ground states, i.e. the left- and right-handed helicoids.
If $\delta D<0$, the ground states are the two catenoids.
In the sine-Gordon description, this term (and other similar terms) gives rise to the cosine terms. 


%

\end{document}